\documentclass{aa}

\usepackage{graphicx}

\usepackage{txfonts}

\usepackage{amsmath}
\usepackage{amssymb}
\usepackage{graphicx}
\usepackage{textcomp}
\usepackage{gensymb}
\usepackage{listings}

\usepackage{subcaption}
\usepackage[switch, modulo]{lineno}
\usepackage{geometry}
\usepackage{multirow}
\usepackage{lscape}
\usepackage{longtable}
\usepackage{booktabs}
\usepackage{pdfpages}
\usepackage[flushleft]{threeparttable}
\usepackage{hyperref}
\usepackage{comment}
\hypersetup{
  colorlinks   = true, 
  urlcolor     = blue,
  linkcolor    = blue, 
  citecolor   = blue 
}

\bibpunct{(}{)}{;}{a}{}{,} 

\begin{document}

\title{Dwarf galaxies in the MATLAS survey: Hubble Space Telescope observations of nuclear star clusters}
\author{M\'elina Poulain\inst{1}, Francine R. Marleau\inst{2}, Pierre-Alain Duc\inst{3}, Rub{\'e}n S{\'a}nchez-Janssen$^{4}$, Patrick R. Durrell$^{5}$, Sanjaya Paudel$^{6}$, Rebecca Habas\inst{7}, Oliver M{\"u}ller$^{8,9,10}$, Sungsoon Lim$^{11,12}$, Nick Heesters$^{8}$, J{\'e}r{\'e}my Fensch$^{13}$}
\titlerunning{HST observations of MATLAS NSCs}
\authorrunning{M. Poulain et al.}

\institute{Space Physics and Astronomy Research Unit, University of Oulu, P.O. Box 3000, FI-90014, Oulu, Finland
\\e-mail: melina.poulain@oulu.fi, melina.poulain45@gmail.com
\and Institute f{\"u}r  Astro- und Teilchenphysik, Universit{\"a}t Innsbruck, Technikerstra{\ss}e 25/8, Innsbruck, A-6020, Austria
\and Universit\'e de Strasbourg, CNRS, Observatoire astronomique de Strasbourg, UMR 7550, F-67000 Strasbourg, France
\and UK Astronomy Technology Centre, Royal Observatory Edinburgh, Blackford Hill, Edinburgh EH9 3HJ, UK
\and Dept. of Physics, Astronomy, Geology, and Environmental Sciences, Youngstown State University, Youngstown, OH 44555 USA
\and Department of Astronomy and Center for Galaxy Evolution Research, Yonsei University, Seoul 03722
\and INAF – Astronomical Observatory of Abruzzo, Via Maggini, 64100 Teramo, Italy
\and Institute of Physics, Laboratory of Astrophysics, École Polytechnique Fédéale de Lausanne (EPFL), 1290 Sauverny, Switzerland
\and Institute of Astronomy, Madingley Rd, Cambridge CB3 0HA, UK
\and Visiting Fellow, Clare Hall, University of Cambridge, Cambridge, UK
\and Division of Science Education, Kangwon National University, Chuncheon, Republic of Korea
\and Department of Astronomy, Yonsei University, 50 Yonsei-ro Seodaemun-gu, Seoul, 03722, Republic of Korea
\and Univ. Lyon, ENS de Lyon, Univ. Lyon 1, CNRS, Centre de Recherche Astrophysique de Lyon, UMR5574, 69007 Lyon, France}

\date{}

\abstract
{
In dwarf galaxies, nuclear star clusters (NSCs) are believed to primarily form from the migration and merger of globular clusters (GCs), with a possible contribution from in situ star-forming activity triggered by gas infall.
We present the study of NSCs in 41 MATLAS survey dwarf galaxies including ultra-diffuse galaxies (UDGs), as part of a large follow-up imaging program with the Hubble Space Telescope (HST) Advanced Camera for Surveys (ACS) using the F606W and F814W filters. The sample is biased toward low surface brightness and large dwarfs, i.e., UDG-like galaxies, and includes two galaxies with a double nucleus; 13 newly identified nucleated dwarfs, thanks to HST's high spatial resolution; and five candidate ultra-compact dwarf progenitors.
We modeled the NSCs with a Sérsic profile and derived their structural properties and photometry.
We find the NSC Sérsic index to increase with the luminosity and stellar mass, while no obvious trend is seen for the effective radius and ellipticity. The faint NSCs tend to have a constant color profile, whereas the bright ones have a bluer center, suggesting that the most massive NSCs in our sample might have experienced a mixed formation scenario, including in situ star formation. A significant portion of our NSCs tend to be more massive than for other galaxy samples of similar stellar mass, which could be due to some dwarfs undergoing tidal disruption or an initial formation of massive NSCs from multiple GC mergers and in situ star-forming activity. More observations of resolved NSCs are needed to be able to infer their formation scenario from the structural properties and photometry in dwarfs.
}


\maketitle

\section{Introduction}
Nuclear star clusters (NSCs) are known to be the densest stellar systems in the Universe. These small compact objects usually have an effective radius within 10 parsec and a stellar mass M$_{\ast}^{\rm NSC}$ from $10^5$ to $10^8$ M$_{\odot}$ \citep{Neumayer2020}. Although they reside in the central region of galaxies, their presence depends on the stellar mass of the host (M$_{\ast}^{\rm gal}$). Known as the nucleation fraction, it reaches a maximum of more than $80\%$ for galaxies with M$_{\ast}^{\rm gal}\sim$10$^9$ M$_{\odot}$ \citep{VandenBergh1986,Ferguson1994,denBrok2014,Ordenes2018,Janssen2019}, while an absence of NSC is observed for more than 90\% of dwarf and high-mass galaxies with M$_{\ast}^{\rm gal}\lesssim10^{6}$ M$_{\odot}$  and M$_{\ast}^{\rm gal}\gtrsim10^{11}$ M$_{\odot}$, respectively (e.g., \citealt{Sandage1985,Faber1997,Cote2006,Antonini2015,Poulain2021,Carlsten2022,Su2022,Marleau2024a}). 

Two main pathways appear to take part in the formation of NSCs. The first is the migration scenario, where GCs form in a galaxy, migrate toward its center due to dynamical friction, and finally merge to form a NSC \citep{Tremaine1975,lotz2001,Bekki2007}. This scenario has been reproduced well by N-body simulations (e.g., \citealt{Antonini2012,Lyubenova2019}), and can explain the formation of NSCs with old metal-poor stellar populations. The second pathway is the in situ scenario \citep{Loose1982}, characterized by the formation of NSCs from star-forming activity caused by infalling gas. While the resulting stellar population tends to be young and metal rich, the in situ scenario is not the only way to produce such young NSCs. In Milky Way (MW)-like galaxies, N-body simulations have shown that NSCs can also be formed by the merging of young  metal-rich clusters \citep{Arca-Sedda2015,Schiavi2021}. The variety of stellar populations of some NSCs, for example M54 in the Sagittarius dwarf, imply a formation based on  the migration and on in situ processes (e.g., \citealt{Sills2019,AlfaroCuello2019,Kacharov2022}). An example of a mixed scenario is the wet migration scenario \citep{Guillard2016}, based on hydrodynamical simulations in dwarf galaxies. This scenario proposes the formation of a GC-like star cluster in a gas-rich dwarf galaxy that contains a gas reservoir. The reservoir  allows the cluster to maintain star-forming activity while it migrates toward the galaxy center, producing a rather young blue NSC. Several recent studies have looked for the main formation channel of NSCs and their results suggest that the migration scenario dominates in galaxies with M$_{\ast}^{\rm gal}$<$10^{9}$ M$_{\odot}$, while the in situ formation prevails in more massive galaxies, and both seem to contribute equally around $10^{9}$ M$_{\odot}$ (e.g., \citealt{Turner2012,denBrok2014,Johnston2020,Fahrion2021,Fahrion2022b,Fahrion2022,Fahrion2024}). 

Together with theoretical works, the formation of NSCs is traditionally studied using three main types of observational data. The first are deep imaging surveys, undertaken mostly via ground-based observations. They are very efficient at unveiling large numbers of new NSCs using photometric data, and they offer detailed statistics about the morphology and local environment of the nucleated galaxies (e.g., \citealt{Reaves1983,Binggeli1985,Ordenes2018,Janssen2019,Habas2020,Carlsten2022,Su2022,Paudel2023}). The second is the analysis of   stellar populations using spectrographs. The Space Telescope Imaging Spectrograph (STIS) of the Hubble Space Telescope (HST) was used by \citet{Rossa2006} to produce the largest spectroscopic survey of NSCs in late-type galaxies (LTGs) to date. Recently, the Multi Unit Spectroscopic Explorer (MUSE) and FOcal Reducer and low dispersion Spectrograph (FORS) instruments on the Very Large telescope (VLT) were chosen to investigate the age, metallicity, and star formation history of NSCs in dwarf galaxies (\citealt{Paudel2011,Johnston2020,Fahrion2021,Fahrion2022b}). Finally, the third is the study of their structural properties with high-resolution space-based observations. HST has frequently been used throughout the years for this kind of work. The luminosity and size of NSCs were first measured in LTGs (see, e.g., \citealt{Carollo1997,Matthew1999,Boeker2002,Balcells2003,Boeker2004,Balcells2007,Georgiev2014}), before targeting early-type galaxies (ETGs) with the HST surveys in the Virgo, Fornax, and Coma clusters \citep{lotz2001,Graham2003,Cote2006,Turner2012,denBrok2014} where more than 200 NSCs were characterized. However, those studies are lacking statistics for dwarf galaxies with M$_{\ast}^{\rm gal}$<$10^{8}$ M$_{\odot}$. More recently, the Early Release Observations from the space telescope Euclid \citep{Marleau2024a} and additional studies with HST (e.g., \citealt{Spengler2017,Pechetti2020,Zanatta2021,Hoyer2023}) have highlighted new NSCs in galaxy clusters and within the Local Volume (LV), especially in dwarf galaxies. 

Related to NSCs are the GCs that also populate the halos of galaxies. Comparing the properties of the GCs to those of the NSCs is important in the context of the migration scenario, especially in dwarf galaxies where this formation path is favored due to their shorter dynamical friction timescales \citep{Hernandez1998,OhLin2000}. Several studies already pointed out correlations between these two types of star clusters. In terms of spatial distribution, a lack of bright GCs was observed in the center of nucleated elliptical dwarfs in the Virgo and Fornax clusters by \citet{lotz2001}. A deficit of GCs at small radii in both nucleated and non-nucleated UDGs was reported by \citet{Marleau2024}, and could be caused by either dynamical friction or the stalling of GCs at the core radius in a core dark matter halo. For dwarf galaxies located in galaxy clusters and in the Local Volume (LV), NSCs and GCs also seem to share a similar occupation fraction, dependant on the stellar mass of the host galaxy \citep{Janssen2019,Carlsten2022}. 

Beyond the formation of NSC, it is also interesting to look for nucleated galaxies undergoing a transition into ultra-compact dwarf galaxies (UCDs). UCDs \citep{Hilker1999,Drinkwater2000,Phillipps2001} are compact stellar systems with an effective radius below 100 pc and a stellar mass above $10^6$ M$_{\odot}$ \citep{Evstigneeva2007,Mieske2008,Misgeld2011}. UCDs have been reported in a wide range of environments, in galaxy clusters, galaxy groups, and around field galaxies (e.g., \citealt{Madrid2010,Chilingarian2011,Norris2011,Faifer2017,Bortoli2020,Liu2020,Saifollahi2021}). In the past two decades a number of scenarios have been put forward to explain the nature of UCDs, and these can broadly be placed into two different formation channels. A fraction of the UCDs are expected to correspond to the high-mass end of the GC population \citep{Kroupa1998,Mieske2002,Mieske2012}, while the most massive UCDs are likely to be NSC remnants of tidally stripped nucleated dwarf galaxies (e.g., \citealt{Bassino1994,Bekki2001,Goerdt2008,Mayes2021,Wang2023}).

Studying dwarf galaxies is important to comprehend the formation and evolution of NSCs. However, only a few observing programs using high spatial resolution imaging exist;  most of them   focus  on galaxy clusters.
The sample of 2210 dwarf galaxies identified in the Mass Assembly of early-Type GaLAxies with their fine Structures (MATLAS) survey offers large statistics beyond the LV, up to a distance of about 45 Mpc, and outside the environment of galaxy clusters \citep{Habas2020}. A photometric study based on the MATLAS deep images from the Canada-France-Hawaii telescope (CFHT) was already undertaken on this sample, and revealed 507 nucleated dwarfs \citep{Poulain2021}. HST observations were conducted for 79 dwarf and ultra-diffuse galaxies (UDGs) from the MATLAS survey \citep{Marleau2024}, 41 of which were nucleated. UDGs are defined as dwarf galaxies with an unusually large effective radius and low central surface brightness, typically R$_{\rm e}$>1.5\,kpc and $\mu_{0,\rm gal}$>24\,mag/arcsec$^{2}$, respectively \citep{vanDokkum2015}. Designed to characterize their GC populations, these high-resolution observations marginally resolve the nuclear region of the galaxies and allow   a more detailed study of the NSCs than previously performed in \citet{Poulain2021}. A first paper on the NSCs by \citet{Poulain2025} focused on the detection of multiple star clusters and stellar tidal tails in a handful of dwarfs. The observed tidal tails were successfully reproduced in N-body simulations, and suggest that they are produced by ongoing star cluster mergers. In this paper we derive the NSCs photometric and structural properties in the 41 dwarf galaxies and UDGs, and discuss their possible formation and evolution pathways.

This work is organized as follows. We introduce the MATLAS survey, as well as the dwarf galaxy sample in Sect. \ref{section:MATLAS_sample}. We present the HST observations in Sect. \ref{section:hst}, and describe the sample of nucleated dwarf galaxies  and UDGs in Sect. \ref{section:nuc_sample}. The method used to derive the properties of the NSCs is detailed in Sect. \ref{section:NSC_method} and their structural and photometric properties are shown in Sect. \ref{section:NSC_props}. We discuss the results in terms of formation of NSCs in dwarf galaxies in Sect. \ref{section:GCs}, and their possible and evolution into UCDs in Sect. \ref{section:UCD}. Finally, we summarize and draw our conclusions in Sect. \ref{section:conclusion}.

\section{The MATLAS dwarf sample}
\label{section:MATLAS_sample}

The galaxies selected for this work were identified in the MATLAS\footnote{\url{http://matlas.astro.unistra.fr/WP/}} survey \citep{Duc2015, Duc2020, Bilek2020}. This survey was designed to study the outermost stellar populations, low surface brightness structures, globular clusters, and dwarf galaxy satellites of ETGs in the context of their mass assembly and the build-up of their scaling relations. The observations were carried out with MegaCam on the CFHT from 2012 to 2015. After being processed by the Elixir-LSB pipeline \citep{MagnierCuillandre2004}, the resulting images are suited for the detection and study of low surface brightness structures, with a surface brightness limit down to $28.5-29$ mag/arcsec$^{2}$ in the $g$-band. The ETGs were selected from the sample of the ATLAS$^{3D}$ project \citep{Cappellari2011}, which is composed of multiwavelength observations of a complete sample of 260 ETGs, numerical simulations, and semi-analytic modeling of galaxy formation. The 260 ETGs are located within a 45 Mpc volume and beyond the LV, with Galactic declinations $|\delta - 29\degree|$, Galactic latitudes above 15\degree, and absolute magnitudes in the $K$-band brighter than $-21.5$. The MATLAS sample is composed of 180 ETGs outside the environment of galaxy clusters, thus avoiding the Virgo cluster ETGs imaged in the Next Generation Virgo Survey (NGVS; \citealt{Ferrarese2012}). We note that this sample also contains 59 late-type galaxies (LTGs) following the same selection criteria as the ETGs, aside from the morphology. The data is divided into 150, 148, and 78 fields in the $g$-, $r$-, and $i$-band, respectively, while the 12 closest ETGs were also observed in the $u$-band. The MATLAS images have a field of view of $1\degree\times 1\degree$ and an average seeing of 0.96\arcsec, 0.84\arcsec, 0.65\arcsec, and 1.03\arcsec\,in the $g$-, $r$-, $i$-, and $u$-band, respectively.

The $g$-band fields were chosen for the detection and identification of dwarf galaxies, due to their more complete sky coverage as compared to the other bands. The selection of the dwarfs (see \citealt{Habas2020} for more details) was based on a visual inspection of the images coupled with a semi-automated method making use of \textsc{source extractor} \citep{Bertin1996}. The final sample contains 2210 galaxies and includes $\sim$73\% of elliptical dwarfs (dEs), $\sim$27\% of irregular dwarfs (dIs), as determined from the CFHT images. Moreover, 23\% of the dwarfs were classified as nucleated. From the spectroscopic redshifts available in the literature and derived heliocentric velocities from HI measurements, \citet{Habas2020} found that about 90\% of the dwarfs with known distance, i.e., $\sim$14\% of the sample, are satellites of one of the ETGs or LTGs from ATLAS$^{3D}$;   80\% are satellites of the central ETGs in the MATLAS fields. Based on these results, it is reasonable to assume that a dwarf without a known distance is a satellite of the targeted ETG in the field it was identified, and thus is located at the same distance. A recent spectroscopic study of 56 MATLAS dwarfs observed with the MUSE instrument on the VLT allowed us to obtain 52 new distances from the derived recessional velocities of the galaxies \citep{heesters2023}, 75\% of which (and 79\% of the dEs) are consistent with the distance of the assumed host ETG. Therefore, for the remainder of this paper, we   make use of the measured distances of the dwarfs, when available, or fix the distance to that of the targeted ETG otherwise.

The photometric and structural properties  of the MATLAS dwarfs, and that of their nuclei, were studied in \citet{Poulain2021}. Follow-up HST observations were performed to study the GC population of selected UDGs from \citet{Marleau2021}. The first observations focused on MATLAS-2019, which has recently become an object of interest due to its large GC population \citep{Mueller2021, Danieli2022}. At least 11 of the GCs have been confirmed by MUSE spectroscopy \citep{Mueller2020}, strengthening the capability of detecting star clusters with HST and selecting them based on their color and concentration index. A more recent work made use of a similar method to investigate the GC population of 74 UDGs \citep{Marleau2024}. These follow-up observations are well suited for a detailed analysis of the NSCs of the nucleated galaxies.

\section{HST observations}
\label{section:hst}
For this study a subsample of 79 of the 2210 dwarf galaxies identified in the MATLAS survey were observed with the HST Advanced Camera for Surveys (ACS) in the Wide Field Channel (WFC) mode. The HST observations were performed during  the Cycle 27 program GO-16082 (PI: O. Müller), as well as the Cycle 28 and 29 snapshot programs SNAP-16257 and SNAP-16711 (PI: F. Marleau). We used the reduced charge transfer efficiency corrected (CTE-corrected) drc drizzled images produced by the STScI standard pipeline, which have a pixel size of 0.05 arcsec/pixel. Each dwarf is centered in one of the charge coupled devices (CCDs) to optimize the background estimate and ensure that no object is lost (e.g., GCs) in the chip gap or outside the field of view. The galaxies were observed in the F606W and F814W filters with an exposure time in each filter of 1030\,s for MATLAS-2019, and 824\,s for the others. We computed the VEGAmag zero-points from the \textsc{python} module \textsc{acszpt}\footnote{\url{https://acstools.readthedocs.io/en/latest/acszpt.html}} made by the ACS Team.
The targets of the follow-up observations were selected to study the GC population of the MATLAS UDGs (e.g., \citealt{Mueller2020,Mueller2021,Marleau2021,Marleau2024,Pfeffer2024,Buzzo2025,Haacke2025}). Thus, our sample is mainly composed of low surface brightness galaxies and large dwarf galaxies (see Fig. 2 in \citealt{Marleau2024}).

\section{The nucleated dwarf sample}
\label{section:nuc_sample}

In \citet{Poulain2021}, we defined a nucleus as a compact source located up to 0.5 effective radius from the photocenter of the dwarf, and that is the brightest compact source within one effective radius. Based on this definition, and after removing one likely foreground star contaminant using the Gaia DR2 catalog \citep{GaiaDR2}, 507 nucleated dwarfs were identified in the MATLAS dwarf galaxy sample. We note, however, that $\sim75\%$ of the nucleus sample are located within a galactocentric distance of 0.1 effective radius, or 0.83\arcsec, a value lower than the observed PSF of the ground-based observation of the MATLAS data. For 16 nucleated dwarfs, up to three compact sources of similar magnitude were located close to the photocenter and were brighter than the surrounding point sources. In these cases, we defined those galaxies as multinucleated.

\begin{figure}
\centering
\begin{subfigure}{\linewidth}
\includegraphics[width=\linewidth]{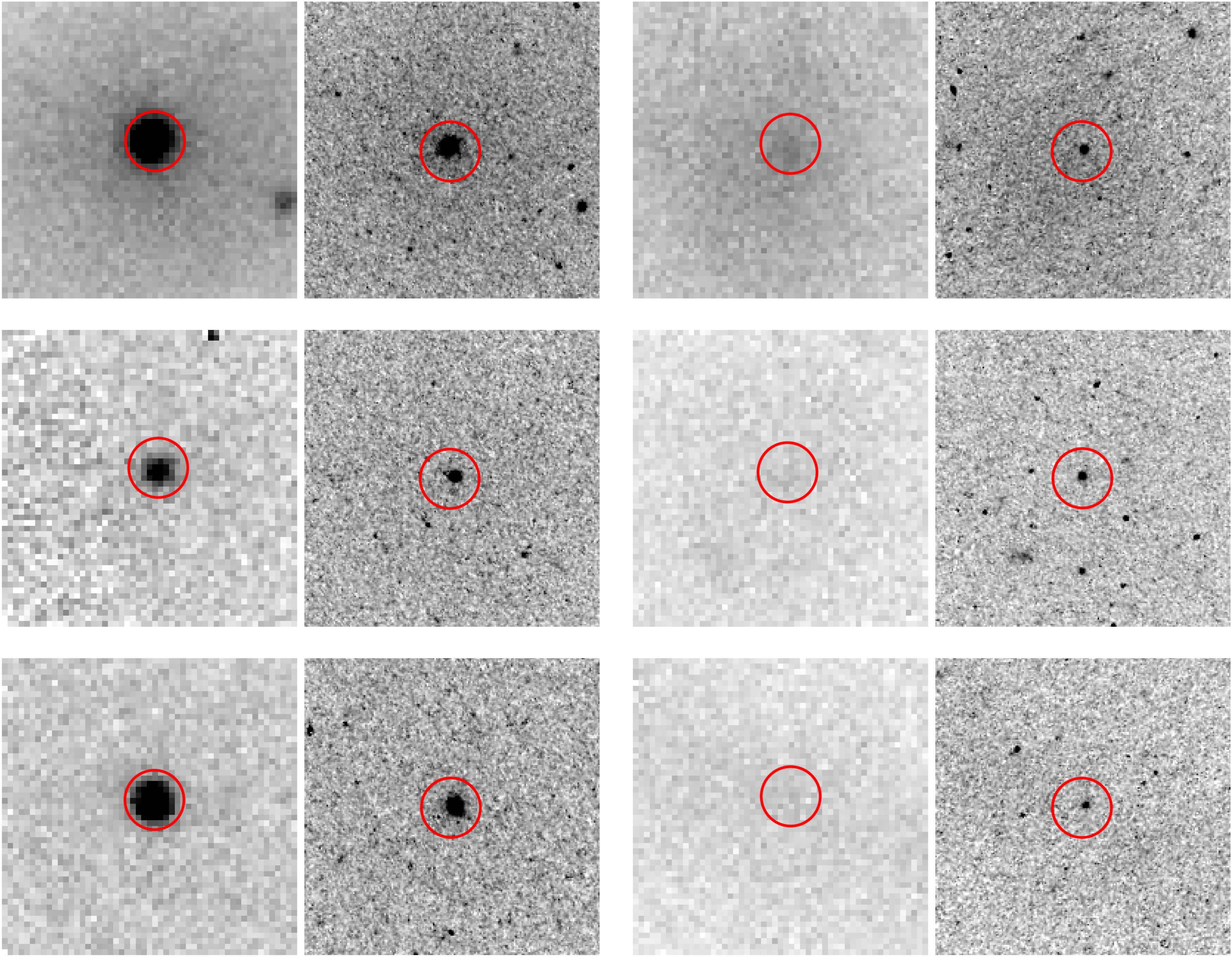}
\end{subfigure}
\caption{Comparison between the CFHT $g$-band (left cutout) and the HST F606W filter (right cutout) observations of the central 10\arcsec$\times$10\arcsec\,area of three nucleated dwarfs from the MATLAS sample (left two columns) and three newly identified (right two columns). We draw an 1\arcsec\,radius circle at the position of the nucleus in each image. North is up and east is to the left.}
\label{fig:compare_matlas_hst}
\end{figure}

\begin{figure*}
\centering
\begin{subfigure}{\linewidth}
\includegraphics[width=\linewidth]{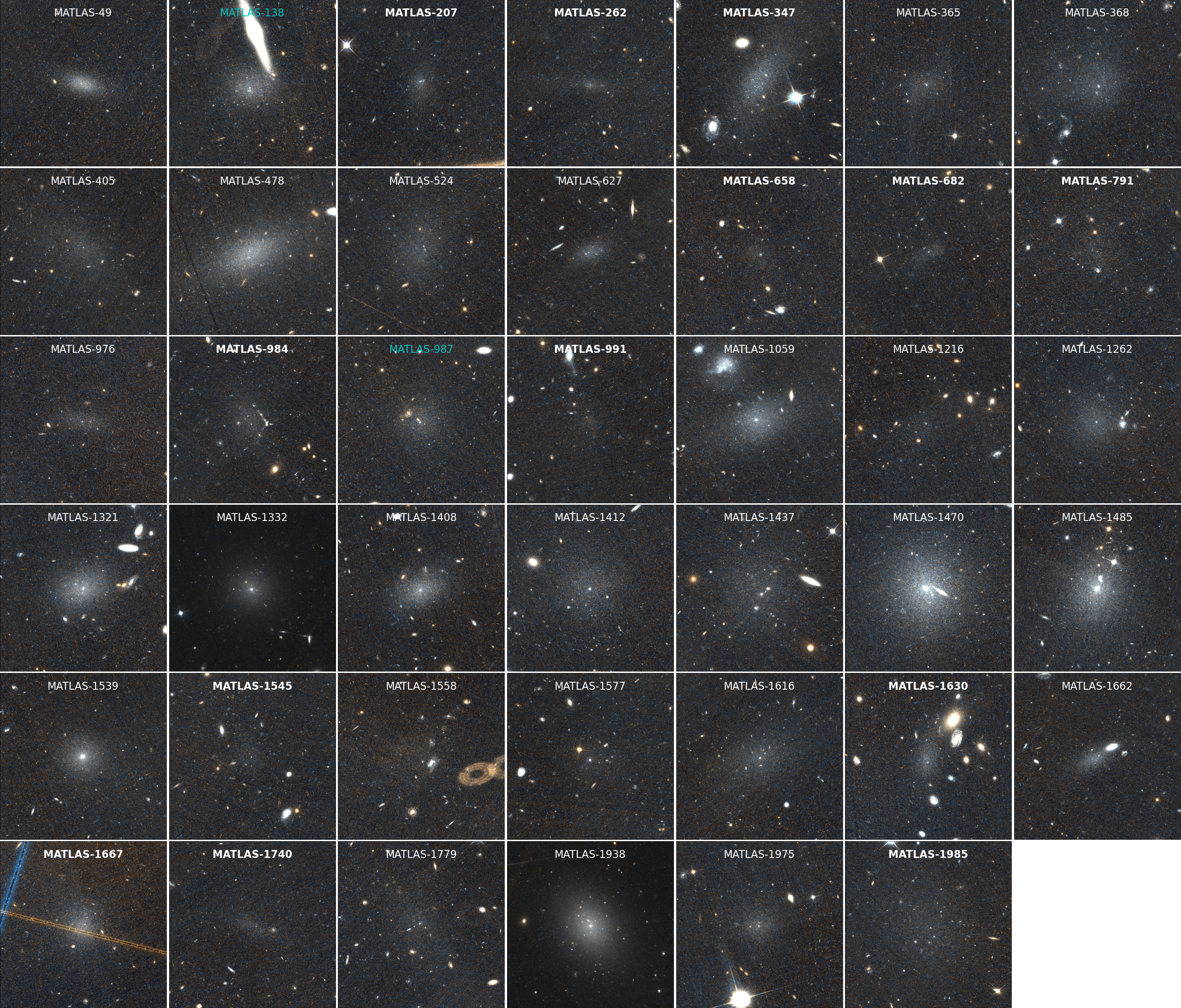}
\end{subfigure}
\caption{HST F606W+F814W color cutouts of the 41 nucleated dwarf galaxies. The names of the  13 new nucleated galaxies are in boldfaced, and those of  the two double-nucleated dwarfs are in cyan. We used a different stretch for MATLAS-1332 and MATLAS-1938 to enhance the visibility of the NSC. Each image is 1\arcmin$\times$1\arcmin. North up and east is to the left.}
\label{fig:colorcutouts}
\end{figure*}

Of the 507 identified MATLAS nucleated dwarfs, 30 have HST observations. However, because the ACS camera has a significantly higher resolution than the MegaCam camera, we were   able to marginally resolve the nuclei, leading to the rejection of two nuclei that were actually background sources, and the identification of 15 new nuclei. Two of these are newly resolved double nuclei located in already flagged nucleated galaxies. This results in a sample of 41 nucleated dwarfs with HST data to be studied. In Fig. \ref{fig:compare_matlas_hst} we show a comparison between the resolution of the central region from the HST and MATLAS images for 6 nucleated dwarfs, including three of these new nuclei;  color cutouts of the 41 nucleated dwarfs observed with HST are displayed in Fig. \ref{fig:colorcutouts}. 
In terms of morphology, 40 of the nucleated dwarfs were classified as dE during the initial classification \citep{Habas2020}, and only one, MATLAS-682, was defined as a dI.

\section{Deriving the properties of the NSCs}
\label{section:NSC_method}

\subsection{S\'ersic modeling}
In order to obtain the structural parameters and photometry of the NSCs, we divided the modeling process into two parts. The aim of the first step was to remove the smooth light profile of the dwarf galaxy from the image. As the background is not constant all over the HST image, we estimated the local background of the galaxy. We computed the median sky from $3\sigma$ clipping in an annulus of radius 15\arcsec\,on the outskirts of the galaxy where the light profile stops decreasing to become constant. To optimize the galaxy modeling, we masked all nuclei with a circle of radius 1\arcsec, a good compromise to cover each NSC without masking a too large region of the galaxies. If necessary, we manually created an additional mask of surrounding bright sources, such as galaxies and stars, with the help of \textsc{saoimageds9} \citep{DS92000}. Finally, we ran \textsc{galfit} \citep{Peng2010} to apply a Sérsic profile \citep{Sersic1963} to the dwarf galaxy, fixing the value of the sky to the estimate. The Sérsic model has long been used in the literature for the NSC modeling (e.g., \citealt{Graham2009,Seth2010,Turner2012,Lyubenova2013,Carson2015,Spengler2017}).
Because of the exposure time of the observations, it can be very difficult to differentiate the observed galaxies from the background given their low surface brightness. As a consequence, the Sérsic models systematically have a smaller R$_{\rm e}$ than the models on the CFHT images \citep{Poulain2021}, as the flux in the outer part of the galaxies blends into the sky noise. We also obtained less accurate photometric measurements, especially for the faintest galaxies. As a consequence, when the galaxies structural properties and photometry were studied, we used the values from \citet{Poulain2021} based on the CFHT images.
The second step focused on the Sérsic modeling of the NSCs. For this, we created a cutout of size 5\arcsec\,centered on the nucleus from the galaxy cleaned image, i.e., the residual image from \textsc{galfit}. The PSF varies with the position on the CCD and the stellar population of the stars; thus, to precisely compute the structural properties of the NSCs, we generated the PSF of each NSC from \textsc{tinytim} \citep{Kirst2011} with a G2V spectral type for the stars, similarly to \citet{Hoyer2023}, and we set the position at the center of the CCD, where the nucleus was observed. It has been shown, however, that a change of spectral type has a mild impact on the NSC modeling \citep{Boeker2004,Georgiev2014}.

The errors on the Sérsic parameters were derived for each reliably modeled NSC by injecting artificial NSCs in the HST residual images. The artificial NSCs were modeled using the \textsc{python} wrapper of the software \textsc{imfit} \citep{Erwin2015} to which we set the parameters of the Sérsic model to be that of the considered NSC. Each artificial source was injected at a random position on the CCD where the target galaxy was centered. In the case of a double nucleated galaxy, we computed the error of each NSC separately. To avoid superposition with bright sources, we isolated the injected NSC from these sources based on a segmentation image from \textsc{source extractor}. We applied the same fitting method to the artificial sources as for the NSCs. To estimate the error on a given parameter, we ran \textsc{galfit} on 100 artificial NSCs leaving only this parameter free to change, and computed the standard deviation of the 100 resulting models.

We obtained a reliable Sérsic model for 14 and 3 NSCs in the F606W and F814W filter, respectively, and for 14 NSCs in both filters. When a model is available for both F606W and F814W images, the parameters tend to agree within the computed errors. Therefore, when applicable, we used the average value of the derived R$_{\rm e,\rm NSC}$, $n$, and ellipticity. All the derived properties and their uncertainties are reported in Table \ref{tab:NSCprop}. Considering the spatial resolution of the observations and the range of distances covered by the host galaxies, the NSCs in the nearest (resp. farthest) galaxies should have R$_{\rm e,\rm NSC}\gtrsim3$\,pc (resp. 10 pc) to start being resolved, increasing the difficulty to marginally resolve the NSCs in the farthest galaxies. Considering galaxies farther than 20\,Mpc, 16 NSCs are close to  unresolved and have effective radius R$_{\rm e,\rm NSC}\lesssim1$\,pix. Those NSCs are highlighted in Table \ref{tab:NSCprop}. In addition, we were unable to derive the structural parameters of the NSC of MATLAS-1630, MATLAS-262, and MATLAS-627, all located at a distance between 30\,Mpc and 46\,Mpc. The remaining galaxies without a NSC model show an intricate central region, consisting of either a possible NSC with stellar tails or multiple bright clusters close the center. The dwarfs with such complex NSCs, including the double nucleated dwarfs, are studied in detail in \citet{Poulain2025}.

\subsection{Color and stellar mass}
We derived the $F606W-F814W$ color of the NSCs from the total magnitudes of the Sérsic modeling when a reliable model was available in both bands. Otherwise, in the case of a one-band-only reliable model, we estimated the color based on aperture magnitude, with a varying aperture size proportional to R$_{\rm e,\rm NSC}$. We measured the aperture magnitudes making use of the \textsc{python} package \textsc{photutils} \citep{Bradley2020}, and applied the aperture corrections reported for the ACS/WFC in \citet{Bohlin2016}. Comparing the obtained colors from aperture and model measurements, we found the best agreement for an aperture radius of 2R$_{\rm e,\rm NSC}$, with a median difference between the two computed colors of 0.04 and a maximum difference below 0.1. For the 12 NSCs without a reliable model, we used the derived photometry from either \citet{Poulain2025} for the complex NSCs or \citet{Marleau2024} for the others, when available.

We converted the magnitudes to the $BVI$ system, applying the same approach as \citet{Hoyer2023}, and employed the synthetic color transformation coefficients of \citet{Sirianni2005}. In the case of a NSC with only one magnitude available, for example $m_{F606W}$, we defined $m_{F814W}$ as $m_{F814W}=m_{F606W}-(m_{\rm F606W, \rm ap}-m_{\rm F814W, \rm ap})$, with $(m_{\rm F606W, \rm ap}-m_{\rm F814W, \rm ap})$ being the color derived from aperture photometry. We corrected the obtained $V$- and $I$-band magnitudes for Galactic extinction using the estimates from \citet{Schlafly2011} at the position of the NSCs.

We computed the stellar mass of the NSCs from the relation between the stellar mass-to-light ratio in the $I$-band $(M_{\ast}^{\rm NSC}/L_{\rm I})$ and the $(V-I)_0$ color defined based on the work and data from \citet{Roediger2015}:
\begin{equation}
    \log_{10}(M_{\ast}^{\rm NSC}/L_{\rm I}) = -0.694+1.335\times(V-I)_0.
\end{equation}

For consistency, we also derived the stellar mass of the dwarfs according to the work of \citet{Roediger2015} using the structural properties and photometry of the dwarfs measured on the MATLAS images published in \citet{Poulain2021} and \citet{Poulain2022}, making use of the stellar mass-to-light ratio in the $g$-band $(M_{\ast}^{\rm gal}/L_{\rm g})$ and the $(g-r)_0$ color relation:
\begin{equation}
    \log_{10}(M_{\ast}^{\rm gal}/L_{\rm g}) = -0.984+2.029\times(g-r)_0.
\end{equation}

On average, we find the obtained stellar masses to be  0.35\,dex less massive than those from \citet{Poulain2021}. This difference is consistent with the findings of \citet{Habas2020}, and is due to the use of the \citet{Bell2003} method to derive the stellar masses in \citet{Poulain2021}, which seems to generate the largest mass differences   compared to other methods \citep{Roediger2015}. Thus, in Sect. \ref{section:Mstar-relation} we   correct the stellar masses derived with \citet{Bell2003} when comparing our galaxies to other samples.

To estimate the uncertainties on the stellar masses, we propagated the errors on the photometry from the artificial sources for the Sérsic models and from \textsc{photutils} for aperture magnitude measurements. We note that, in addition to these errors, scatters in the relations between the mass-to-light ratios and colors will add an average of 0.3 dex uncertainty on the estimated stellar mass \citep{Roediger2015}.

\begin{figure*}
\centering
\begin{subfigure}{\linewidth}
\includegraphics[width=\linewidth]{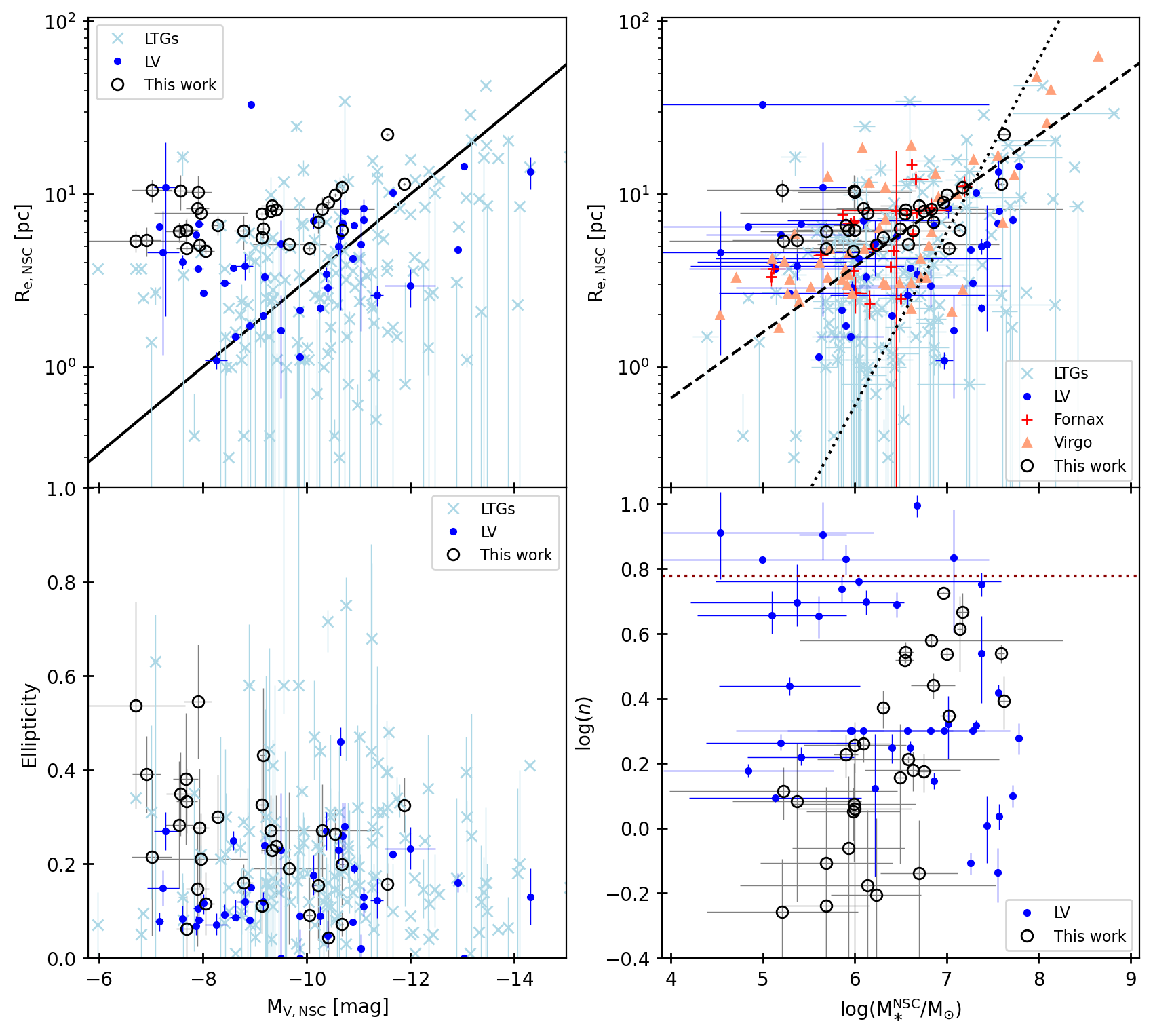}
\end{subfigure}
\caption{Scaling relations between the structural properties of the MATLAS NSCs studied in this work (black open circles)   compared to NSCs from the LV (blue dots),   Virgo (orange triangles), and Fornax (red crosses) clusters, and located in LTGs (light blue crosses), using the structural parameter-based sample described in Sect. \ref{section:NSCcatalogs}. Top: Effective radius as a function of the absolute magnitude in the $V$-band (left) and the stellar mass (right). We show the relation log\,R$_{\rm e,\rm NSC}$=$-2.0-0.25$M$_{\rm V,\rm NSC}$ fitted by \citet{Georgiev2014} (black line), R$_{\rm e} \propto$ M$_{\ast}^{0.38}$ (black dashed line) found by \citet{Bekki2004}, and R$_{\rm e} \propto$ M$_{\ast}$ (black dotted line) suggested by \citet{Turner2012}. Bottom left: Ellipticity vs. the absolute magnitude in the $V$-band. Bottom right: Relation between the Sérsic index and the stellar mass. The dotted line indicates $n=6$.}
\label{fig:struct_prop}
\end{figure*}

\section{Structural and photometric properties}
\label{section:NSC_props}
In this section we present the structural properties of the 31 NSCs with a reliable Sérsic model, as well as the photometry of 41 NSCs. We first compare them to NSC properties from the literature, and then detail their color profile.

\subsection{Catalogs of NSCs}
\label{section:NSCcatalogs}
For comparison purposes, we compiled the results of several NSC studies targeting both dwarfs and more massive galaxies located in various environments. 
We used two types of samples:  structural parameter-based and  photometry-based. The structural  parameter-based sample gathers the properties of NSCs observed by HST and is used to study their R$_{\rm e,\rm NSC}$, ellipticity, and Sérsic index as a function of the NSCs absolute magnitude and stellar mass.
In the LV, \citet{Pechetti2020}\footnote{We only considered the best-quality NSC models (0 and 1) to avoid galaxies highly contaminated by dust.} and \citet{Hoyer2023} recently presented Sérsic modeling of NSCs in galaxies with stellar masses between $10^{6.5}$ and $10^{11}$\,M$_{\odot}$. In the dense environment of galaxy clusters, a first analysis of the size and luminosity of NSCs in elliptical galaxies was performed based on the HST data from the ACS Fornax and Virgo cluster surveys \citep{Turner2012,Cote2006}. We also compare our NSCs to those located in field LTGs at distances up to 40\,Mpc \citep{Georgiev2014,Georgiev2016}.

The type of model fitted to the underlying galaxy light profile impacts the reliability of the derived NSC properties, especially in massive galaxies, as a single Sérsic profile is sometimes used instead of a multicomponent surface brightness profile in the presence of a bulge, bar, or disk \citep{Peng2002}. Because of the use of a single-Sérsic or a core-Sérsic \citep{Graham2003} profile to model the nucleated galaxies, we cleaned up the samples of the ACS Fornax and Virgo cluster surveys from objects that would have required a multicomponent fitting. In the case of \citet{Cote2006}, we removed the galaxies with disks or bars identified in \citet{Ferrarese2006}, while we removed the galaxies from \citet{Turner2012} brighter than a total blue magnitude of 13.5\,mag or described as better fitted by a multi-Sérsic profile.

The photometry-based sample is composed of the measured photometry and stellar masses of the NSCs and their host galaxy. It combines the NSCs from the structural parameter-based sample to ground-based observations. We added the NSC population of dwarf galaxies surrounding LTGs from \citet{Carlsten2022}, as well as the photometric measurements and derived stellar masses from \citet{Janssen2019} and \citet{Spengler2017} in the Virgo cluster, and from \citet{Ordenes2018} and \citet{Su2022} in Fornax.

\subsection{Relations between structural properties}
\label{section:NSC_props_relations}
In Fig. \ref{fig:struct_prop} we present several relations between the structural properties of the NSCs. We first investigated the variation of the R$_{\rm e,\rm NSC}$ with M$_{\rm V,NSC}$ and the stellar mass $M_{\ast}^{\rm NSC}$. Considering  only studies where NSCs were modeled, we compared the MATLAS NSCs to previously presented samples. When the stellar mass was not available, we used the available photometry and mass-to-light ratios from \citet{Roediger2015} to calculate it ourselves. 
Several studies showed that NSCs with $M_{\ast}^{\rm NSC}\gtrsim10^{6}$\,M$_{\odot}$ or M$_{\rm V,NSC}\lesssim-9\,$mag tend to be larger as they get brighter and more massive (e.g., \citealt{Georgiev2014,Norris2014,Neumayer2020,Hoyer2023}), and that this trend appears to break for fainter, less massive NSCs. We show the relation fitted by \citet{Georgiev2014} in the top left panel and see that, while the brightest MATLAS NSCs with M$_{\rm V,NSC}\lesssim-10\,$mag seem to follow the relation, we do not see an effect of the luminosity or mass on R$_{\rm e,\rm NSC}$ when considering the whole sample. We show in the top right panel of Fig. \ref{fig:struct_prop} the relation R$_{\rm e,\rm NSC} \propto$ M$_{\ast}^{0.38}$ found for star clusters formed from GC mergers in the simulations from \citet{Bekki2004}. \citet{Turner2012} pointed out that the relation should evolve depending on the number of mergers that the NSC undergoes, with the most massive ones following a steeper relation such that R$_{\rm e,\rm NSC} \propto$ M$_{\ast}$, which we also added to the plot. While some NSCs from our sample might follow either of these two relations, we cannot confirm the assumption that these NSCs formed from the migration scenario based on their R$_{\rm e,\rm NSC}$, mass, and luminosity only.

In the bottom left panel we investigate a potential relation between the ellipticity and the stellar mass of the NSCs. \citet{Spengler2017} and \citet{Hoyer2023} found that the brighter  and more massive the NSCs, respectively, the flatter they are. However, we do not see such a trend for the MATLAS dwarfs, LTGs, and combined LV samples. On the contrary, the bright MATLAS NSCs tend to be rounder than the faint ones, but given the large errors on the ellipticity, we cannot affirm any relation.

Finally, we show the Sérsic index $n$ as a function of the stellar mass. In the LV, a weak anti-correlation was reported and fitted by \citet{Pechetti2020}, in agreement with the structural properties found for bright NSCs in nearby LTGs by \citet{Carson2015}. Adding data for NSCs in dwarf galaxies, \citet{Hoyer2023} performed another regression resulting in a less steep slope. However, this difference of relation might  be driven by the high $n$ values as \citet{Pechetti2020} considered only NSCs with $n<6$, while \citet{Hoyer2023} used the whole sample. We indicate this cut in the plot.
Unlike the findings in the LV, the massive MATLAS NSCs seem to have a larger Sérsic index than the low-mass ones. This correlation is consistent with the observations of \citet{Spengler2017} for NSCs in the Virgo cluster. However, we can see that the MATLAS NSCs fit into the large scatter found for the LV population. Thus, considering both the LV and MATLAS samples, it is hard to find any relation.
In terms of stellar distributions, a higher Sérsic index means a population that becomes more concentrated toward the center of the NSC within R$_{\rm e,\rm NSC}$ when $n<6$; \citet{Carson2015} reported that Sérsic profiles with $n>6$ indicate a difference in the extent of the profile beyond R$_{\rm e,\rm NSC}$, rather than a difference in central stellar concentration. 

\begin{figure}
\centering
\begin{subfigure}{\linewidth}
\includegraphics[width=\linewidth]{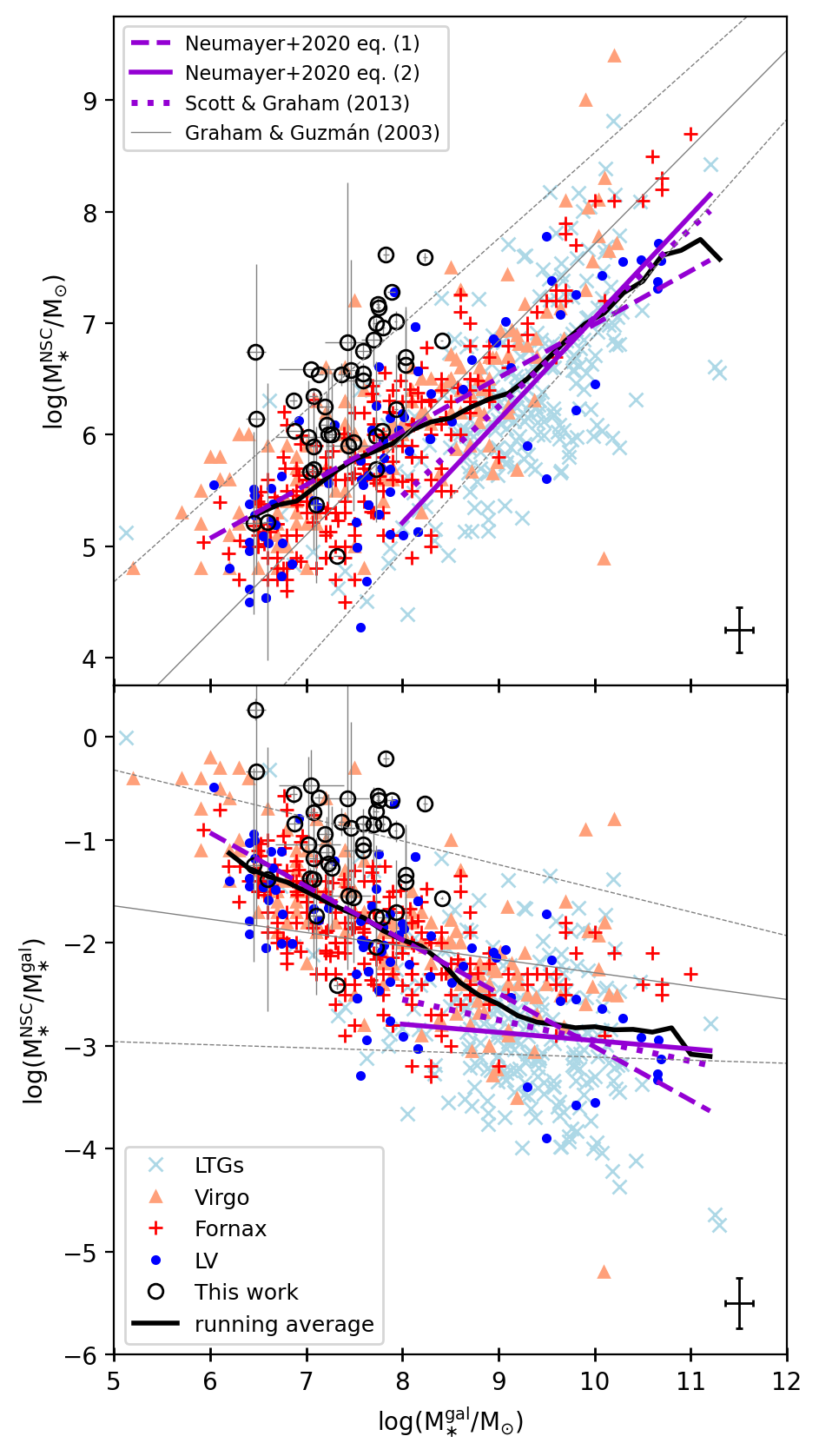}
\end{subfigure}
\caption{NSC stellar mass to galaxy stellar mass relation (top) and NSC-to-galaxy mass ratio vs. galaxy stellar mass (bottom) for the MATLAS NSCs from this work (black open circles), LV (blue dots), Virgo (orange triangles), Fornax (red crosses), and LTG (light blue crosses) photometry-based sample from Sect. \ref{section:NSCcatalogs}. We computed the running average of log M$_{\ast}^{\rm NSC}$ in bins with a width of 1 dex and a moving step size of 0.2 dex in log M$_{\ast}^{\rm gal}$ considering all the NSCs (black solid line). In the bottom plot, we compare the running average to  Eqs. 1 (purple dashed line) and 2 (purple line) of \citet{Neumayer2020}, as well as the relation found for ETGs in \citet{Scott2013} (purple dotted line) and dEs in \citet{Graham2003} (gray line; with dashed lines for the errors). The error bars (lower right corner) represent the median uncertainties.}
\label{fig:Mstar_gal_NSC}
\end{figure}

\subsection{NSC-to-galaxy stellar mass relation}
\label{section:Mstar-relation}

Several studies have reported correlations between some properties of the NSCs and the host galaxies, especially their luminosity and mass, in  dwarf galaxies and in more massive galaxies (e.g., \citealt{Balcells2003,Graham2003,Boeker2004,Turner2012,Scott2013,Poulain2021,Carlsten2022,Hoyer2023}). In this section we focus on the NSC-to-galaxy mass ratio. This relation has been studied for a large range of galaxies and environments, and thus is well suited for a comparison with our sample. We show at the top of Fig. \ref{fig:Mstar_gal_NSC} the relation between the stellar mass of the NSC and that of its host galaxy, while we plot the NSC-to-galaxy mass ratio as a function of the galaxy stellar mass in the bottom part. We compare the HST NSCs to those observed in the LV,  Fornax, and Virgo clusters, and in LTGs, using all the samples with available stellar masses for both the NSCs and hosts described in Sect. \ref{section:NSCcatalogs}. 

In agreement with the literature, we observe an overall increase in the mass of the NSC with that of the host. We studied this increase by computing the running average of the NSC mass per bin of galaxy stellar mass, combining all the samples. Interestingly, we find that a majority of the NSCs in the HST sample fall above the running average, suggesting that the NSCs tend to be more massive in our sample. While these NSCs appear to also deviate from the relation fitted by \citet{Neumayer2020} on dwarf galaxies, considering the error bars on the MATLAS dwarfs, they mostly agree within the estimate uncertainties on the relation fitted on dEs from the Coma cluster \citep{Graham2003}. We investigated the deviant galaxies by looking in detail at the properties of the dwarf galaxies with high $M_{\ast}^{\rm NSC}$ and mass ratio in the Virgo cluster. We find that either they tend to have a very high nuclear-to-total luminosity fraction or they are UDGs. According to \citet{Wang2023}, all have properties of UCD progenitors and are  undergoing tidal disruption. The possible evolution of the MATLAS nucleated galaxies observed with HST into UCDs is discussed in Sect. \ref{section:UCD}. Another possible explanation could be that some of these galaxies might have experienced more GC mergers than other dwarfs due to a larger population of GCs, leading to more massive NSCs \citep{Turner2012}. Over the past decade, UDGs with a large number of GCs have been identified (e.g., \citealt{Lim2020,Mueller2021,forbes2024,Marleau2024}).

We note that we observe an overall change in the relation at a galaxy stellar mass of about $10^{9} M_{\odot}$ in the lower panel of Fig. \ref{fig:Mstar_gal_NSC}, already mentioned in previous works \citep{Ordenes2018,Janssen2019,Neumayer2020,Su2022}. The running average of low-mass galaxies follows the fitted relation $M_{\ast}^{\rm NSC}\propto \sqrt{M_{\ast}^{\rm gal}}$ of \citet{Neumayer2020}, while the massive galaxies have an almost flat relation, similar to the one fitted by \citet{Neumayer2020} using the massive galaxies with dynamical and spectroscopic measurements for the NSC mass from \citet{Erwin2012}, and the relation fitted for ETGs and dEs in \citet{Scott2013}. Part of this change in the relation could be due to a change in the dominant morphology of the galaxies: ETG for the low-mass sample and LTG for the high-mass sample. The effect of the morphology on the NSC-to-galaxy stellar mass relation has been investigated in several studies \citep{Georgiev2016,Neumayer2020}. In addition, \citet{Scott2013} reported a steepening of the relation when the sample contains a nuclear disk. While the presence of such a disk was not reported in the samples used here, it might explain the deviation from the relation of some galaxies with $M_{\ast}^{\rm gal}>10^{10} M_{\odot}$.

\subsection{Color of the NSCs}
\label{section:color}
In this section we study the stellar populations of the NSCs by looking at their integrated $(F606W-F814W)_0$ color as well as the variation of the color with radius, i.e., their color profile. The modeled NSCs show a range of color indices from 0.54 to 1.23 with a mean  $(F606W-F814W)_0=0.78\pm0.14$ mag, similar to the observed colors of GCs in the MATLAS UDGs \citep{Mueller2021,Marleau2024}. No trends are observed between the color of the NSCs and its absolute magnitude, the galaxy color, or the galaxy stellar mass.

\begin{figure}
\centering
\begin{subfigure}{\linewidth}
\includegraphics[width=\linewidth]{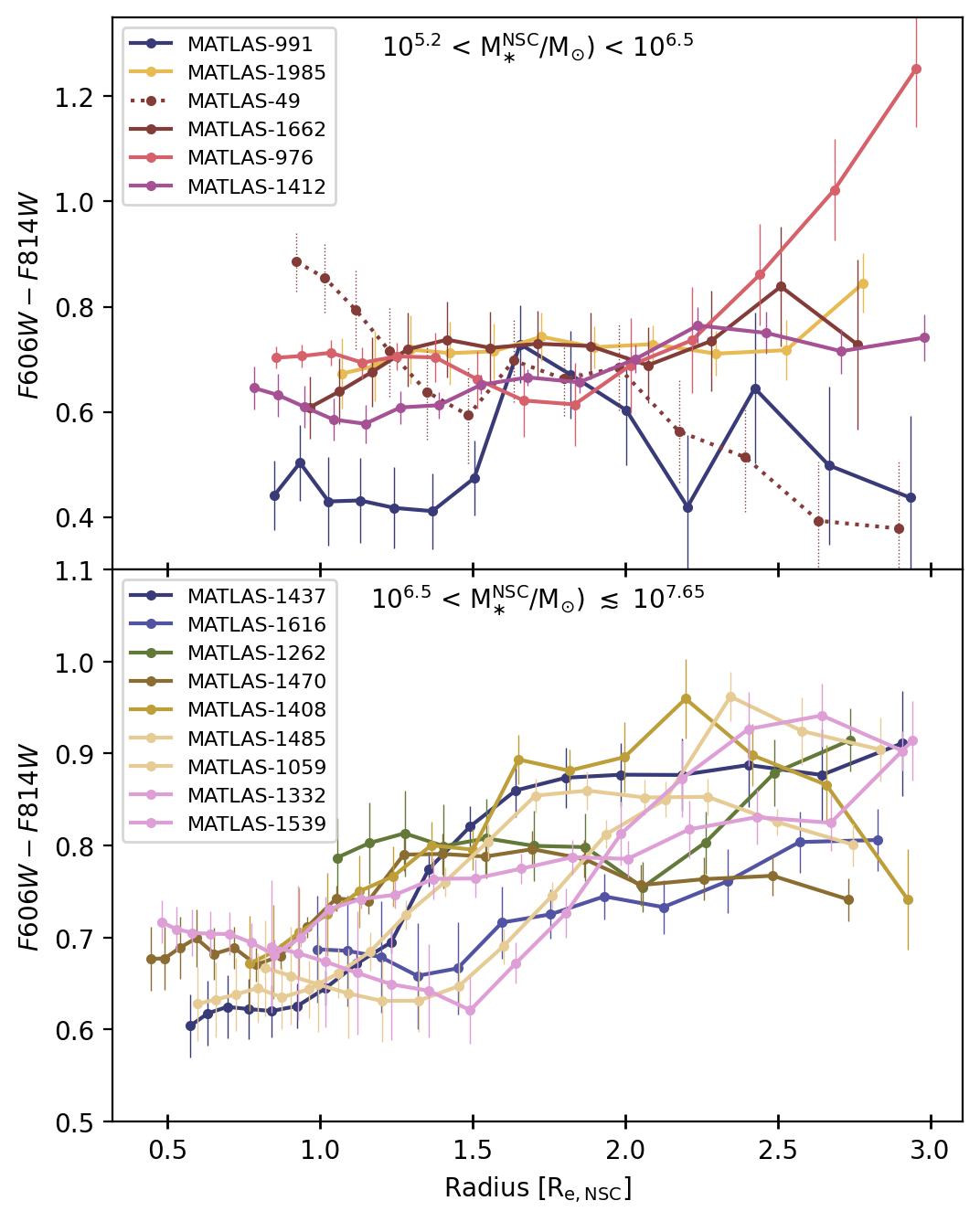}
\end{subfigure}
\caption{Color profiles observed within 3R$_{\rm e,\rm NSC}$ for the 15 modeled NSCs with R$_{\rm e,\rm NSC}$>1\,pix. The NSCs are order by increasing M$_{\ast}^{\rm NSC}$ and split into two subplots: M$_{\ast}^{\rm NSC}<10^{6.5}$ M$_{\odot}$ (top) mostly showing constant profiles, and M$_{\ast}^{\rm NSC}>10^{6.5}$ M$_{\odot}$ (center) with varying profiles from redder outskirts to a bluer center. The peculiar profile, from bluer outskirts to an inner redder color, of MATLAS-49 is highlighted with a dotted line.}
\label{fig:expl_color}
\end{figure}

We defined the color profiles of the NSCs as the difference between the light profiles in the F606W and F814W images that we produced making use of \textsc{autoprof} \citep{Stone2021b} fixing the same center and ellipses in both bands. We selected only the 15 marginally resolved NSCs with R$_{\rm e,\rm NSC}$>1\,pix to ensure there were multiple pixels. We display the color profiles within 3R$_{\rm e,\rm NSC}$ ordered by increasing M$_{\ast}^{\rm NSC}$ in Fig. \ref{fig:expl_color}, and the individual profiles in Fig. \ref{fig:color-profiles}, to investigate a possible change in the profile with the NSC mass. We observe that a large majority of the low-mass NSCs, with M$_{\ast}^{\rm NSC}\lesssim10^{6.5}$ M$_{\odot}$, tend to have an almost constant profile, while the massive ones show a change in color:  a red outer area and an inner region that gets bluer toward the photocenter. We note that both types of profile can be found at M$_{\ast}^{\rm NSC}$ from $10^{6}$ to $10^7$ M$_{\odot}$.

The absence of a color gradient in the center of the low-mass NSCs means there is a homogeneous stellar population, arguing in favor of GC merging. On the contrary, the change in color can be explained by a change in stellar population. At the high-mass end of NSCs in dwarf galaxies, a combination of formation scenarios has been reported \citep{Fahrion2021,Fahrion2022b,Fahrion2022}. This composite scenario matches our observations, starting with GC infall producing the outer region red color,  followed by in situ star formation causing a younger bluer stellar population at the center. Given the low spatial resolution of the NSCs, the observed trend needs to be confirmed by observing more NSCs, especially in nearby galaxies.

An exception to this color trend is the low-mass NSC of MATLAS-49 with a $\log(M_{\ast}^{\rm NSC}/M_{\odot})$ of 5.91. This NSC has an opposite profile to the massive NSCs, with a blue stellar population in the outer region that gets redder toward the photocenter. In the context of dwarf galaxies, this type of profile could be due to circumnuclear star formation, as observed for some NSCs in dEs \citep{Paudel2020} following the wet migration scenario predictions. According to this scenario, MATLAS-49 should be a gas-rich dE, but no HI-detection was reported for this galaxy in \citet{Poulain2022}. Given its location, we estimate its HI mass to be less than $5\times10^7\,$M$_{\odot}$.

\section{Formation of NSCs from GCs}

\label{section:GCs}
\begin{figure}
\centering
\begin{subfigure}{\linewidth}
\includegraphics[width=\linewidth]{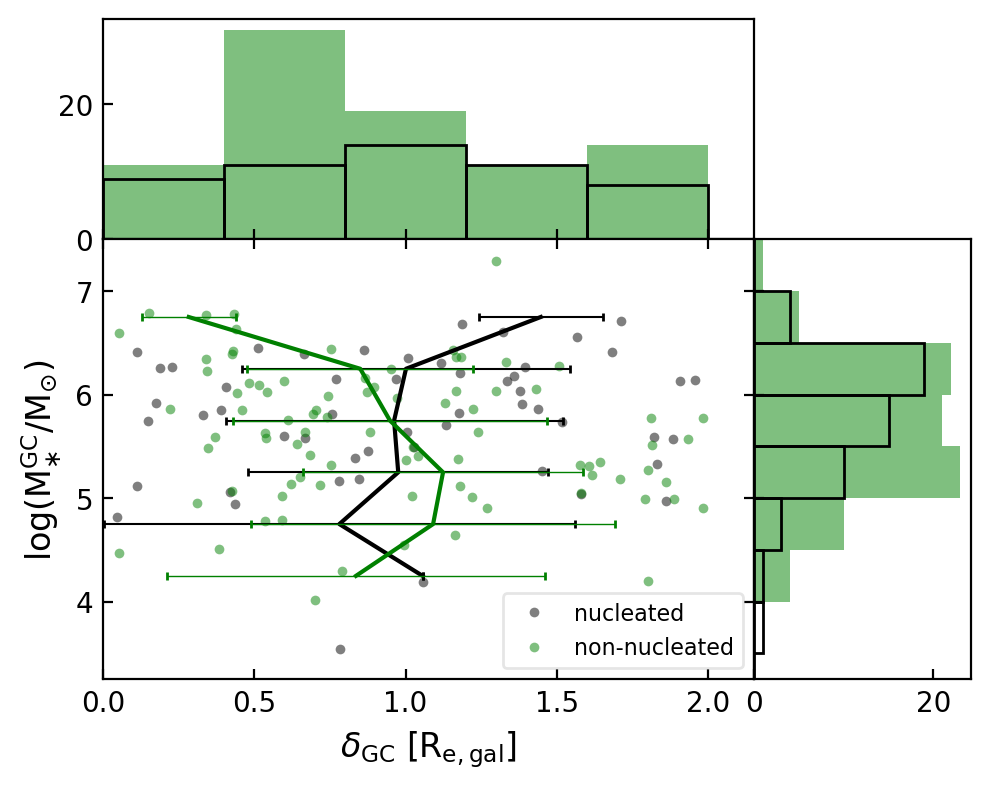}
\end{subfigure}
\caption{M$_{\ast}^{\rm GC}$ as a function of the radial distribution of the GCs for the nucleated (black) and non-nucleated (green) dwarfs. We show the running average of the radial distance $\delta_{GC}$ per bin 0.5\,log(M$_{\ast}^{\rm NSC}$/M$_{\odot}$) with lines. The error bars represent the standard deviation.}
\label{fig:Radial_mass}
\end{figure}

The GC population of the MATLAS UDGs observed by HST is studied in detail in \citet{Marleau2024}. 
Making use of this work, we first investigated the radial distribution of the GCs as a function of their stellar mass to question the effect of dynamical friction. In a second part, we look at the R$_{\rm e}$\,--\,luminosity relation of  the GCs and the NSCs in the context of the migration scenario.

\subsection{GC radial distribution}

The strength of the dynamical friction depends on the steepness of the density profile of the galaxy, but also on the mass of the GC and its initial position \citep{Hernandez1998, lotz2001,Arca-Sedda2014,Arca-Sedda2015}, so that the most massive GCs, and GCs at small radial distances, are the fastest to migrate to the center and are the most likely to merge to form a NSC, and galaxies with the steepest profiles have the shortest orbital decay timescales. To explore this GC mass segregation, we look at the radial distance of the GCs of the nucleated and non-nucleated dwarfs as a function of their mass. We only modeled the marginally resolved GC candidates detected in the 17 dwarfs located below 25 Mpc, as the GC systems are dominated by unresolved sources at greater distances (see \citealt{Marleau2024} for further details). Following the same approach to the NSCs described in Sect. \ref{section:NSC_method}, we obtained a reliable Sérsic model for 142 GCs in at least one of the two filters, from which we derived the colors and stellar masses. The properties of the GCs are available in Table \ref{tab:GCprop}, and we indicate the GCs with R$_{\rm e,\rm GC}\lesssim1$\,pix, as for the NSCs.

In Fig. \ref{fig:Radial_mass} we show the dependence of the radial distribution on the stellar mass of the GCs within 2R$_{\rm e,\rm gal}$ of the nucleated and non-nucleated dwarfs, together with the running average of the radial distance $\delta_{GC}$ per bin, 0.5\,log(M$_{\ast}^{\rm GC}$/M$_{\odot}$). We looked for an effect of the Sérsic index of the host galaxy on the radial distribution of the GCs using the structural properties from \citet{Poulain2021}. We find that the nucleated galaxies have, on average, a higher Sérsic index than the non-nucleated dwarfs with 1.01 and 0.85, respectively, suggesting a steeper density profile for the nucleated dwarfs and a more efficient dynamical friction. However, the galaxies show a wide range of Sérsic index, from about 0.5 to 1.6, regardless of the presence of a NSC, and we do not find any influence of the Sérsic index on the radial distribution of the GCs.
We observe that GCs in both nucleated and non-nucleated galaxies have a similar radial distribution for M$_{\ast}^{\rm GC}$<\,10$^{6.5}\,$M$_{\odot}$, with the GCs being located at all $\delta_{GC}$ --as shown by the large standard deviation on the running average-- except for GCs with M$_{\ast}^{\rm GC}$>\,10$^{6}\,$M$_{\odot}$, which are not found beyond 1.5\,R$_{\rm e,\rm gal}$ in the non-nucleated galaxies. We note that the sample is  incomplete at the low-mass end since we consider here only marginally resolved GCs. Thus, the radial distribution of GCs with M$_{\ast}^{\rm GC}$<\,10$^{5}\,$M$_{\odot}$ cannot be assessed. Considering the most massive GCs with M$_{\ast}^{\rm GC}$>\,10$^{6.5}\,$M$_{\odot}$, we find opposite trends: the GCs are mostly found within 0.5\,R$_{\rm e,\rm gal}$ in non-nucleated dwarfs, while they are located above 1\,R$_{\rm e,\rm gal}$ in nucleated dwarfs. These tendencies are based on low statistics, but might highlight the effect of dynamical friction, where the most massive GCs have migrated to the center of the non-nucleated dwarf, while the most massive GCs in the nucleated might have merged to form the NSCs. Hence, the lack of these GCs in the central region is similar to the findings of \citet{lotz2001} for nucleated dEs in the Virgo cluster.

\begin{figure}
\centering
\begin{subfigure}{\linewidth}
\includegraphics[width=\linewidth]{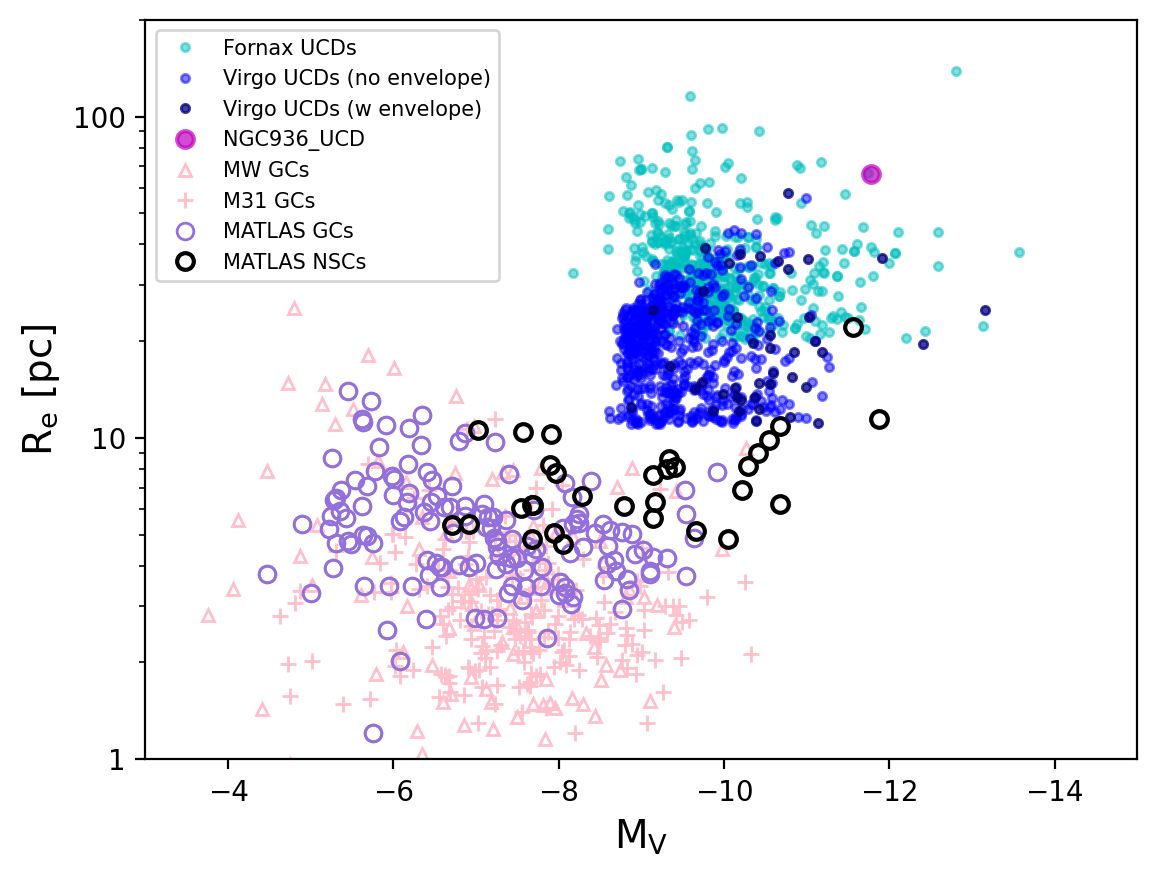}
\end{subfigure}
\caption{Scaling relation between R$_{\rm e}$ and M$_V$ for the MATLAS NSCs (black open circles) and GCs (purple open circles) compared to the GC population of the MW (pink triangles; \citealt{Harris1996}, 2010 edition) and M31 (pink crosses; \citealt{Peacock2010}), the UCDs in the Virgo with (dark blue dots) and without (blue dots) an identified outer envelope (\citealt{Liu2020}), and Fornax (cyan dots; \citealt{Saifollahi2021}), and the forming UCD around NGC 936 (magenta dot; \citealt{Paudel2023}).}
\label{fig:ScalingRelation}
\end{figure}

\subsection{R$_{\rm e}$ -- luminosity relation}

The simulations of NSC formation from GC mergers undertaken in \citet{Bekki2004} have shown that the relationship between the structural properties and the luminosity of the NSCs differ from those found for GCs. We investigated in Sect. \ref{section:NSC_props_relations} the relation between R$_{\rm e}$ and M$_V$ of the MATLAS NSCs. We now compare this scaling relation to that of the MATLAS GCs. In Fig. \ref{fig:ScalingRelation} we plot R$_{\rm e}$ as a function of M$_V$ for both types of star clusters, and compare the MATLAS GCs to those in the MW citep{Harris1996}, 2010 edition) and M31 \citep{Peacock2010}. Unsurprisingly, the MATLAS GCs occupy a similar range of M$_V$ to the MW and M31 GCs. In agreement with \citet{Bekki2004}, we observe a different trend for the GCs than for the NSCs, where the maximum R$_{\rm e}$ reached by the GCs at a given M$_V$ decreases toward brighter GCs, while it is constant for the NSCs. 
\citet{Hoyer2023} noted that the NSC outliers in the LV seem to follow the size-mass and size-luminosity relations observed for the GCs. More statistics for low-mass NSCs is needed to assess whether they  fall into the GC regime.

\begin{figure}
\centering
\begin{subfigure}{\linewidth}
\includegraphics[width=\linewidth]{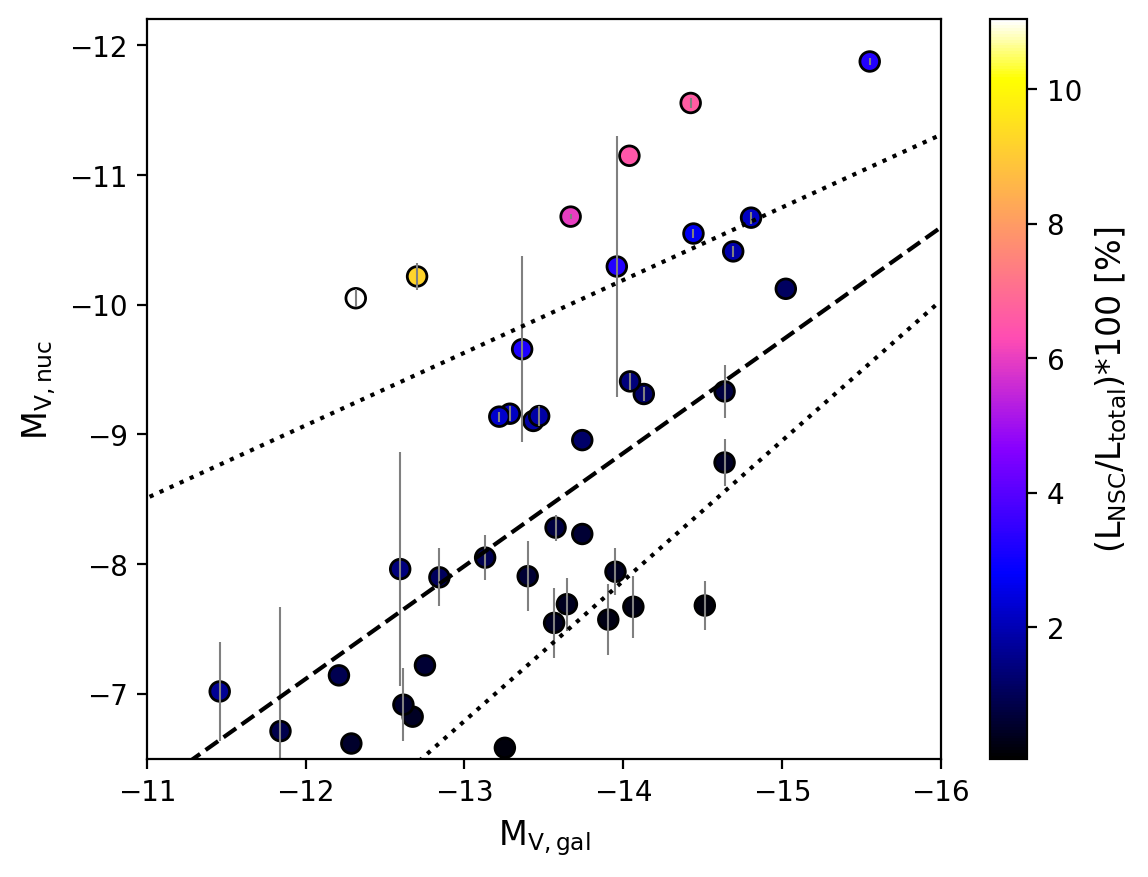}
\end{subfigure}
\caption{NSCs M$_V$ to host galaxy M$_V$ relation. The color bar indicates the nuclear-to-total-luminosity fraction. We show the NSC-to-galaxy luminosity relation fitted by \citet{Graham2003} (dashed line) together with the estimated upper and lower errors on the relation (dotted lines).}
\label{fig:Luminosity_fraction}
\end{figure}

\section{Nucleated dwarfs as UCD progenitors}
\label{section:UCD}

Ultra-compact dwarf galaxies are a class of galaxies as bright as dwarf galaxies but more compact, in between the size of dwarfs and GCs. One of their formation scenarios implies that UCDs formed from the stripped nuclei remnants of tidally disrupted nucleated dwarf galaxies (e.g., \citealt{Zinnecker1988,Freeman1993,Bekki2001}). Simulations argue that this is the predominant formation path of the most massive UCDs \citep{Pfeffer2014,Mayes2021}. Moreover, this scenario is also supported by the observation of tidal features around some UCDs \citep{Voggel2016,Schweizer2018}, asymmetric or elongated shapes \citep{Wittmann2016}, diffuse halos \citep{Evstigneeva2008,Liu2020}, similarities between the stellar populations of UCDs and NSCs \citep{Paudel2010,Norris2015,Janz2016,Wang2023}, and the presence of MBHs in UCDs \citep{Graham2020,Mayes2024}. An example of a UCD candidate in the process of formation has already been identified in the MATLAS dwarf galaxy sample. MATLAS-167 is located at the end of a tidal tail extending from the giant S0 galaxy NGC936; a detailed analysis of the object can be found in \citet{Paudel2023}.

We investigated whether some of the NSCs in our HST sample could be UCD progenitors. In Fig. \ref{fig:ScalingRelation} we compare the R$_{\rm e}$ -- luminosity scaling relation of the modeled MATLAS NSCs to the UCDs from the Virgo and Fornax galaxy clusters \citep{Liu2020,Saifollahi2021}. We split the Virgo sample to highlight the UCDs with identified outer envelopes, and we only show UCDs with R$_{\rm e}>20$\,pc in the Fornax sample, as \citet{Saifollahi2021} question the reliability of the measurement of smaller sizes. During the formation of UCDs, up to 98\% of the stellar mass of the MATLAS dwarfs will get stripped \citep{Bekki2001,Wang2023}, leaving the NSCs slightly larger and brighter than they currently are. As the MATLAS NSCs appear to form a bridge between the GCs and UCDs, we suggest that they could be UCD progenitors if their dwarf galaxy host undergoes tidal disruption.

\begin{figure*}
\centering
\begin{subfigure}{\linewidth}
\includegraphics[width=0.33\linewidth]{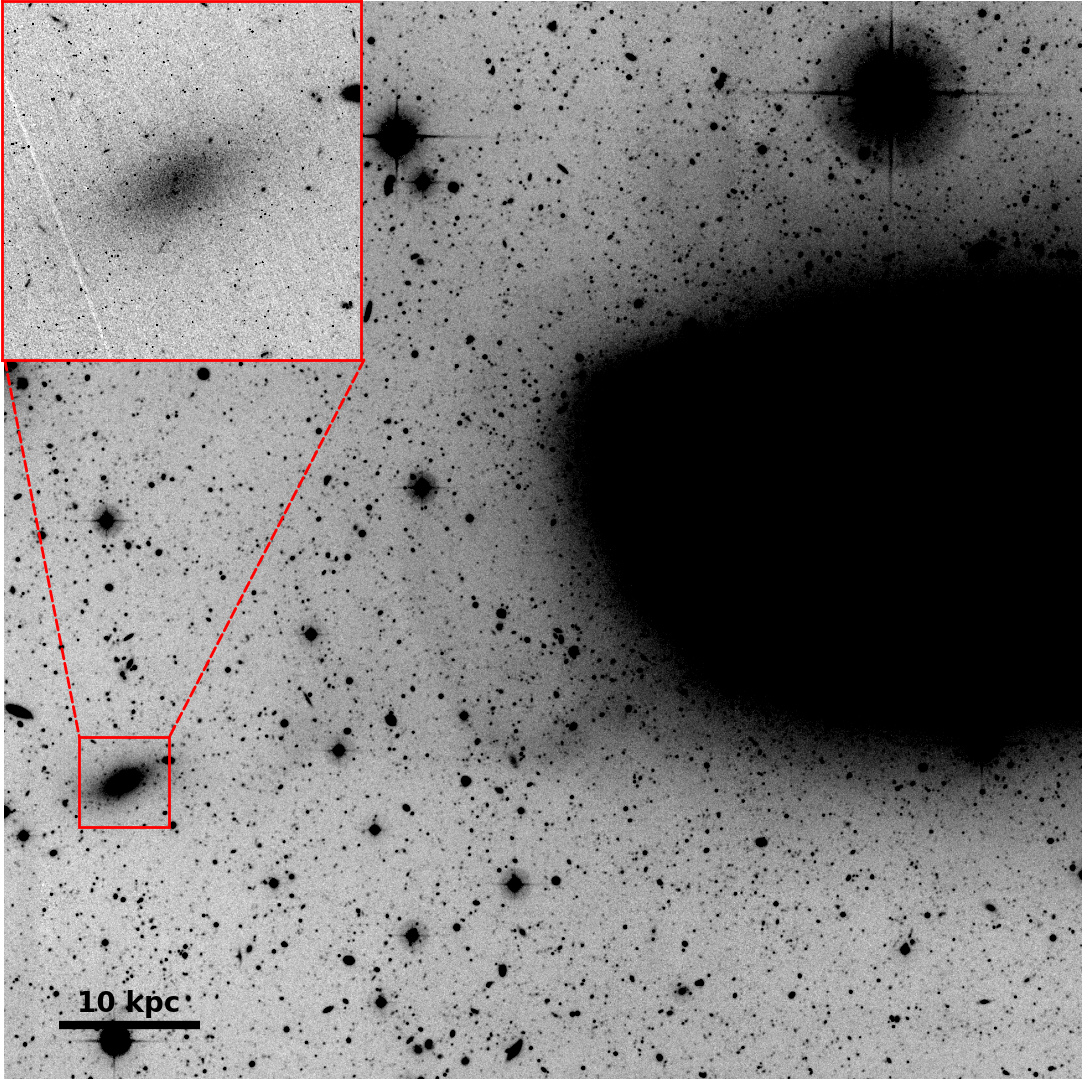}
\includegraphics[width=0.33\linewidth]{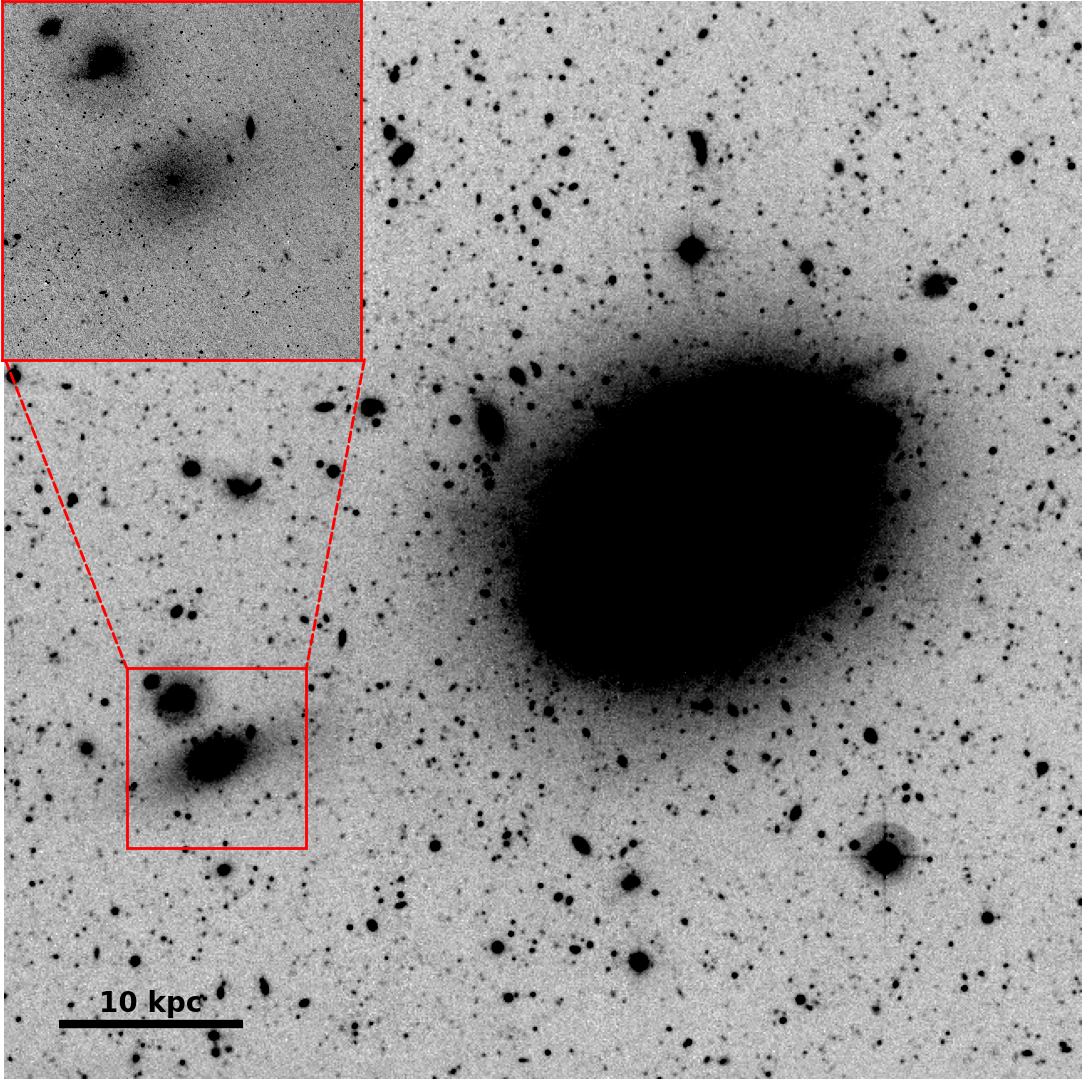}
\includegraphics[width=0.33\linewidth]{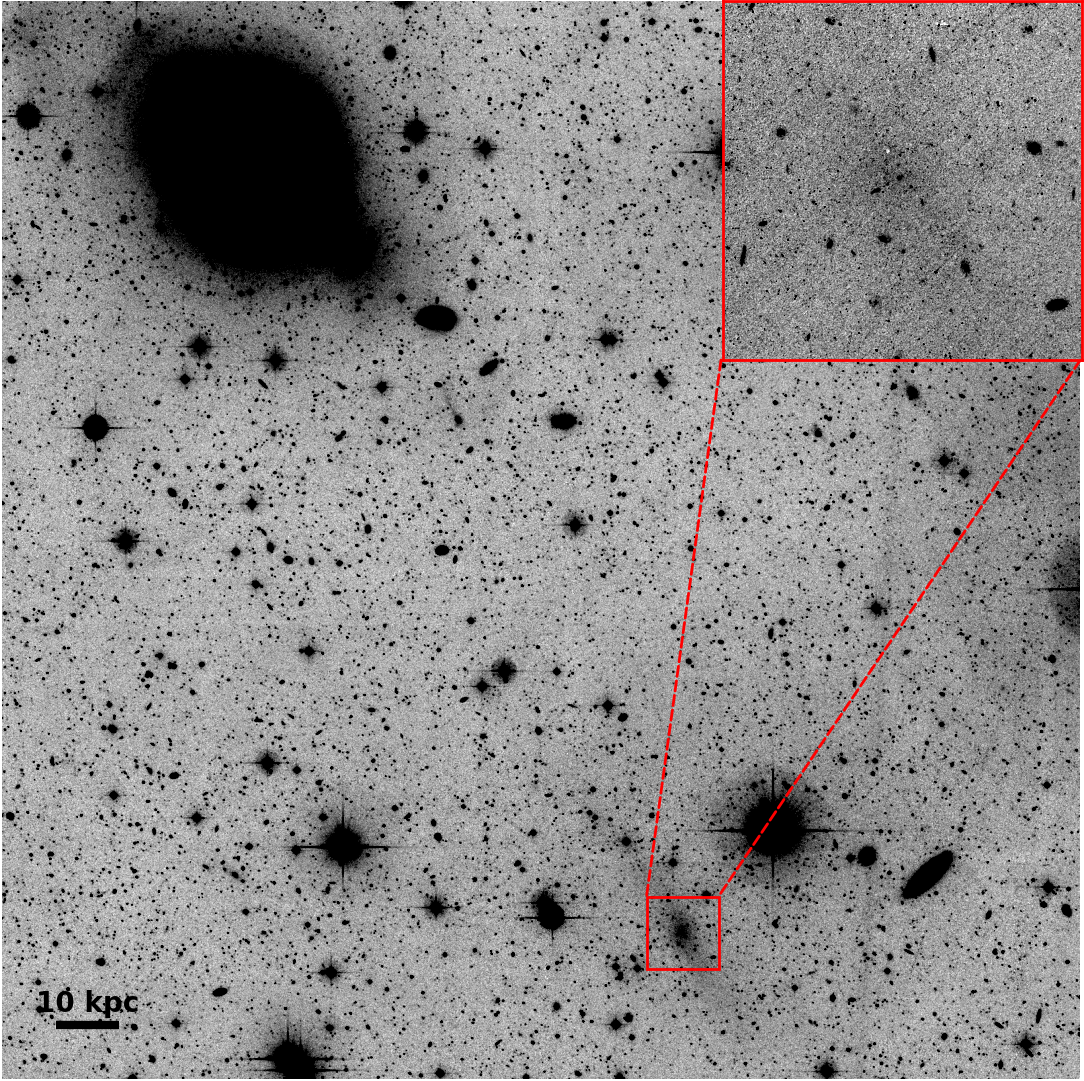}
\includegraphics[width=0.33\linewidth]{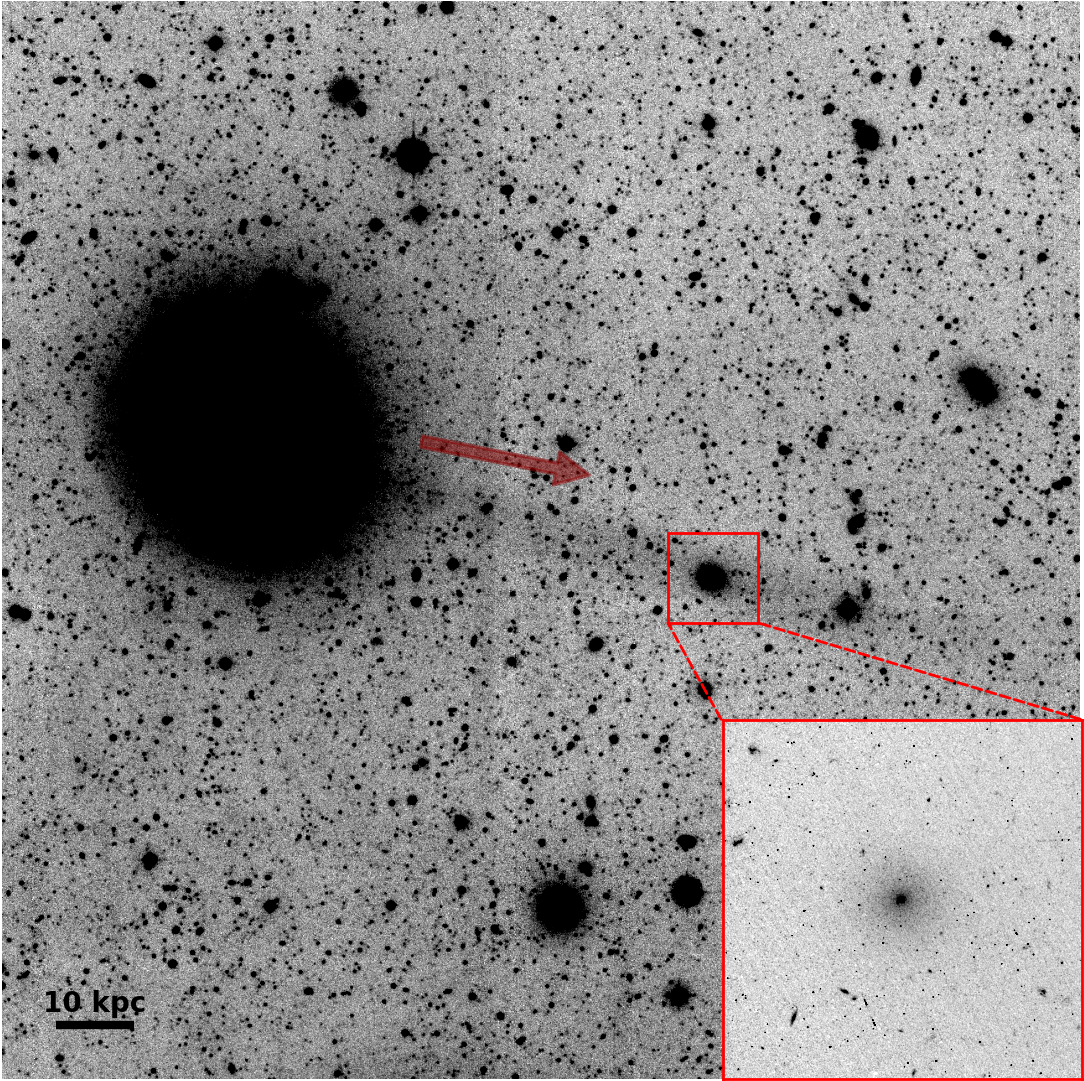}
\includegraphics[width=0.33\linewidth]{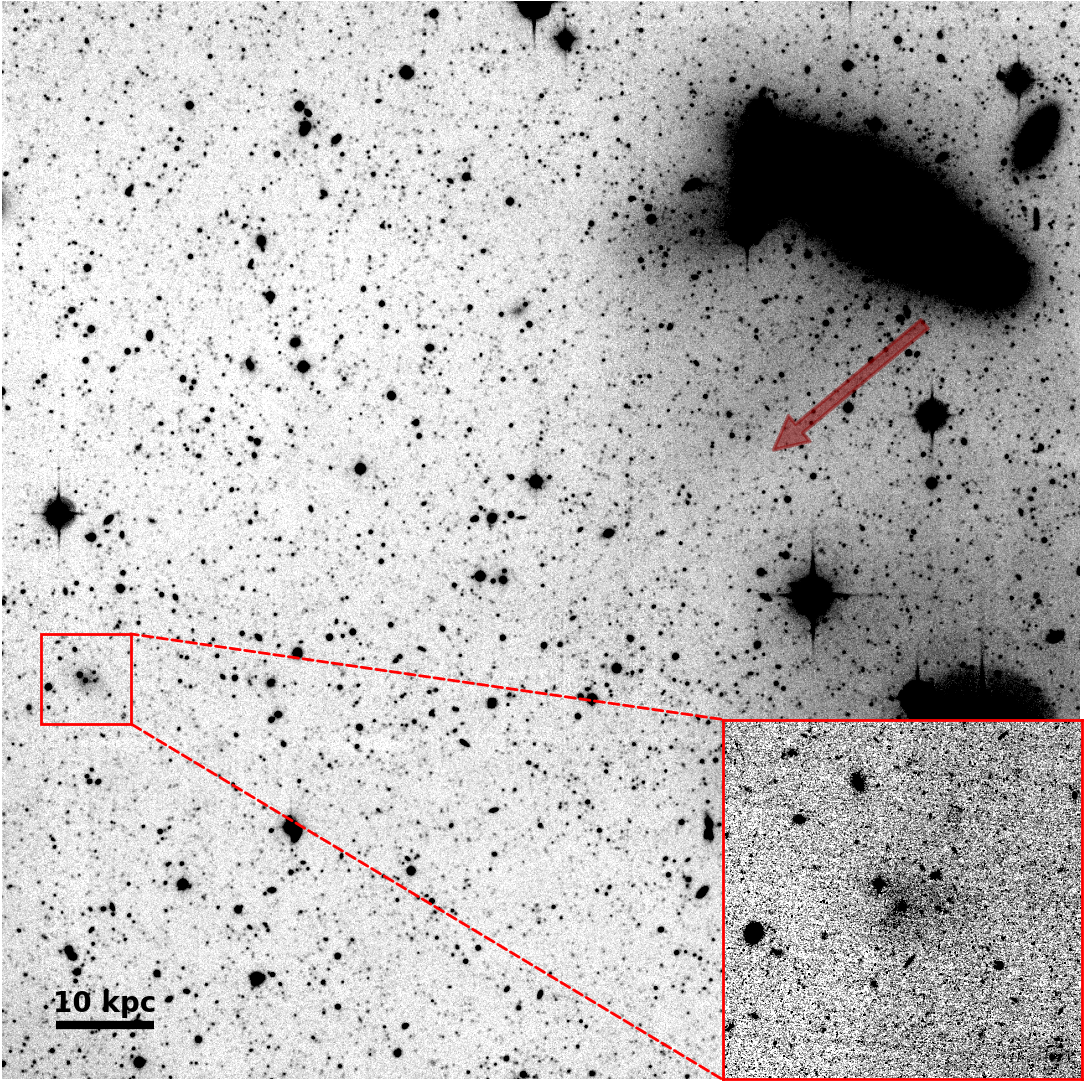}
\end{subfigure}
\caption{Five candidate UCD progenitors in our HST sample. Top row: Three tidally elongated UDGs, MATLAS-478, MATLAS-1059, and MATLAS-1779 (from left to right), located near a massive galaxy and showing a low nuclear-to-total luminosity fraction of 0.2\%, 2.2\%, and 3.3\%, respectively. Bottom row: Dwarf galaxy MATLAS-1539 (left) located on top of a tidal tail of NGC5198 (red arrow), and the dwarf galaxy MATLAS-1577 (right) with a disturbed morphology extended toward a stellar stream from NGC5308 (red arrow). Both dwarfs have a nuclear-to-total luminosity fraction larger than 6\%. All cutouts from the MATLAS CFHT $g$-band image are $12\arcmin\times12\arcmin$, except for MATLAS-1059 ($6\arcmin\times6\arcmin$) and MATLAS-1779 ($15\arcmin\times15\arcmin$). We show $1\arcmin\times1\arcmin$ zooms into the dwarfs and UDGs from the HST F606W observations.}
\label{fig:UCD_progenitors}
\end{figure*}

The study of \citet{Wang2023} presents different phases of UCD formation from tidally stripped nucleated dEs in the Virgo cluster. They suggest that, during the first transitional phase, tidal stripping can remove more than 90\% of the galaxy stellar mass, creating strongly nucleated dwarfs with a nuclear-to-total-luminosity fraction reaching $8\% - 35\%$. In Fig. \ref{fig:Luminosity_fraction} we show the NSC-to-galaxy luminosity relation of our HST sample;  the nuclear-to-total luminosity fraction of the NSCs is shown as a color bar. The NSCs show an overall good agreement with the relation derived by \citet{Graham2003}. We investigated whether the five dwarfs outliers with a fraction $\gtrsim 6 \%$ show signs of tidal disruption. We identified tidal features from nearby massive galaxies related to two of them, MATLAS-1539 and MATLAS-1577, which have a fraction of 6.6\% and 9.2\%, respectively. The dwarfs and corresponding tidal features are visible in Fig. \ref{fig:UCD_progenitors}. MATLAS-1539 seems compact and is located on top of a tidal tail from NGC5198, while MATLAS-1577 has a disturbed morphology and is extended toward a stellar stream from NGC5308. We note that the NSC of MATLAS-1539 is the largest of the modeled ones, with a R$_{\rm e,\rm NSC}$ of 21.58\,pc, and already follows a similar M$_V$--R$_{\rm e}$ relation to the UCDs (see Fig. \ref{fig:ScalingRelation}). Following these results, even though only MATLAS-1577 meets the criteria of strongly nucleated dwarfs from \citet{Wang2023}, we suggest that  MATLAS-1539 and MATLAS-1577 are both candidate UCD progenitors. Among the galaxies with a nuclear-to-total luminosity fraction below 6\%, we found three tidally elongated UDGs located close to massive galaxies. According to \citet{Wang2023}, at the early steps of tidal disruption, nucleated dEs fall into the regime of UDGs and show tidal features such as an extended stellar envelope or tidal tails. Based on this statement, we add these three UDGs to the list of candidates, and show them in Fig. \ref{fig:UCD_progenitors}.

\section{Summary and conclusions}
\label{section:conclusion}
In this work we investigated the formation and evolution of NSCs by studying their structural and photometric properties in a dwarf galaxy sample biased toward low surface brightness and diffuse (UDG-like) galaxies located beyond the LV and outside the environment of galaxy clusters. We made use of HST follow-up observations of 79 MATLAS galaxies with the ACS using the F606W and F814W filters. This sample is composed of 41 nucleated dwarfs, 13 of which are newly identified thanks to HST high spatial resolution images, and two have a double nucleus. 

We modeled the NSCs with a Sérsic profile and obtained a reliable result for 31 of them; the derived parameters are provided in Table \ref{tab:NSCprop}. We compared their structural properties and photometry to NSCs in nearby field LTGs, and in dwarf galaxies from the LV, and from the Virgo and Fornax galaxy clusters. The R$_{\rm e,\rm NSC}$ of the MATLAS NSCs does not tend to be larger as they get brighter and more massive. Moreover, while we do not see an influence of the absolute magnitude on the ellipticity, the Sérsic index of the MATLAS NSCs gets larger with M$_{\ast}^{\rm NSC}$, as observed in the Virgo cluster.

The relation between the NSCs and their host galaxy shows  that a non-negligible fraction of the MATLAS NSCs tend to be brighter and more massive than NSCs in typical dwarf galaxies. This could be  due to the galaxies undergoing tidal disruption or to an increased growth of NSCs in UDGs, thus producing more massive NSCs.

We derived the color profiles within 3R$_{\rm e,\rm NSC}$ of the modeled NSCs and remark that, while the low-mass NSCs tend to have an almost constant profile consistent with the migration scenario, the most massive NSCs have a steep slope, from an outer redder color to an inner bluer color, suggesting a change in stellar population, which might be caused by a contribution from in situ star formation. 

Following the work of \citet{Marleau2024} on the GC systems of the HST MATLAS UDGs, we studied the distribution and properties of the GCs in the context of the migration scenario. 
We modeled the GCs in the MATLAS dwarfs up to 25 Mpc and find that, although with low statistics, the most massive GCs with M$_{\ast}^{\rm GC}$>\,10$^{6.5}\,$M$_{\odot}$ are distributed oppositely in the nucleated and non-nucleated dwarfs: the former seem to lack massive GCs in their central 0.5R$_{\rm e,\rm gal}$ region, while they are located within 0.5R$_{\rm e,\rm gal}$ for the latter. Comparing the R$_{\rm e}$ -- luminosity scaling relation of the GCs to the NSCs, we find a different relation for the two types of star clusters, in agreement with \citet{Bekki2004}.

The most massive UCDs are believed to be the NSC remnants of tidally stripped nucleated dwarfs. In that context, we searched for UCD progenitors among the 41 NSCs with available photometry. We found that the NSCs appear to link the GC and UCD populations in the R$_{\rm e}$ -- luminosity scaling relation, suggesting that some of them could become UCDs if their dwarf host undergo tidal disruption. Looking for strongly nucleated dwarfs, we identified two UCD progenitor candidates, MATLAS-1539 and MATLAS-1577, that have a nuclear-to-total-luminosity fraction of 6.7\% and 8.7\%, respectively, and show signs of a possible tidal interaction with their assumed host ETG. For the galaxies with a low nuclear-to-total-luminosity fraction, we found three tidally elongated UDGs, likely in the early stage of tidal disruption, which we added to our list of UCD progenitor candidates.

Overall, the observed trends need to be confirmed by studying even larger samples of nucleated dwarfs using a deep and high-resolution imaging facility (e.g., the Euclid Wide Survey; \citealt{Scaramella2022}), combined with spectroscopic follow-up observations to trace the NSC star formation history.

\begin{acknowledgements}
The authors would like to thank the referee for the insightful comments, which helped to improve the manuscript. This research is based on observations from the NASA/ESA Hubble Space Telescope obtained at the Space Telescope Science Institute, which is operated by the Association of Universities for Research in Astronomy, Incorporated, under NASA contract NAS5-26555. Support for Program number GO-16257 and GO-16711 was provided through a grant from the STScI under NASA contract NAS5-26555. This research was supported by the International Space Science Institute (ISSI) in Bern, through ISSI International Team project No. 534. This research made use of AutoProf, a package for galaxy image photometry \citep{Stone2021b}. M.P. is supported by the Academy of Finland grant No. 347089. O.M. and N.H. are grateful to the Swiss National Science Foundation for financial support under the grant number PZ00P2\_202104. P.R.D. gratefully acknowledges support from grant HST-GO-16257.002-A. S.P. acknowledges support from the Mid-career Researcher Program (No. RS-2023-00208957). S.L. acknowledges the support from the Sejong Science Fellowship Program by the National Research Foundation of Korea (NRF) grant funded by the Korea government (MSIT) (No. NRF-2021R1C1C2006790). RH acknowledges funding from the Italian INAF Large Grant 12 - 2022.
\end{acknowledgements}

\bibliographystyle{aa}
\bibliography{biblio.bib}

\onecolumn
\begin{appendix}

\section{NSC properties}
\label{AppendixA}

Table \ref{tab:NSCprop} presents the structural and photometric properties of the 43 NSCs identified in the 41 nucleated dwarf galaxies observed with HST. For each NSC, the table includes the host galaxy ID, distance and stellar mass (M$_{*}^{\rm gal}$), together with the NSC RA/Dec coordinates, extinction corrected absolute magnitude in the $V$-band (M$_{\rm V,NSC}$), $(V-I)_0$ color, effective radius (R$_{\rm e,\rm NSC}$), Sérsic index $n$, ellipticity, and stellar mass (M$_{*}^{\rm NSC}$). Two values of M$_{*}^{\rm gal}$ are shown: the one derived for this work based on \citet{Roediger2015} stellar mass-to-light ratios, and in parentheses, the one used in either \citet{Poulain2021} or \citet{Poulain2022} based on \citet{Bell2003} method. The last column indicates the band from which the parameters were derived. If the parameters are based on one band, 2R$_{\rm e,\rm NSC}$ aperture photometry was applied in both bands to derive the color. When both bands were used, R$_{\rm e,\rm NSC}$, $n$, and the ellipticity correspond to the average value of both bands. An absence of band means that there is no reliable model for the NSC, and that the derived photometry was picked from either \citet{Poulain2025} or \citet{Marleau2024}, when available. The NSCs are sorted by increasing RA. MATLAS-138 and MATLAS-987 have a double nucleus, and thus have two lines.

\setlength{\tabcolsep}{5pt}
\begin{table*}
\caption{\label{tab:NSCprop}Structural and photometric properties of the 43 NSCs.}
\resizebox{\linewidth}{!}{
\begin{tabular}{cccccccccccc}
\toprule
Host galaxy & Dist &  $\log(\frac{M_{*}^{\rm gal}}{M_{\odot}})$ & RA & Dec & M$_{\rm V,NSC}$ & $(V-I)_0$ & R$_{\rm e,\rm NSC}$ & $n$ & Ellipticity & $\log(\frac{M_{*}^{\rm NSC}}{M_{\odot}})$ & Band\\
  & [Mpc] &  & [deg] & [deg] & [mag] & [mag] & [pc] &  &  &  & \\
\toprule
MATLAS-49 & $35.9$ & $7.73\pm0.02\,(8.00)$ & $20.7647$ & $9.1018$ & $-7.57\pm0.27$ & $1.01\pm0.69$ & $10.45\pm2.38$ & $1.19\pm0.40$ & $0.35\pm0.09$ & $5.99\pm0.67$ & F606W\\
MATLAS-138 & $37.5$ & $8.04\pm0.03\,(8.27)$ & $27.8990$ & $22.2933$ & $-8.78\pm0.18$ & $1.10\pm0.55$ & $6.15^*\pm1.38$ & $1.51\pm0.21$ & $0.16\pm0.04$ & $6.63\pm0.52$ & F606W\\
 &  &  & $27.8991$ & $22.2935$ & $-9.33\pm0.21$ & $1.01\pm0.42$ & $8.61^*\pm0.36$ & $0.73\pm0.33$ & $0.23\pm0.00$ & $6.70\pm0.42$ & F606W\\
MATLAS-207 & $35.3$ & $6.88\pm0.08\,(7.24)$ & $42.4459$ & $-1.1773$ & $-7.22\pm0.06$ & $1.11\pm0.07$ & - & - & - & $6.03\pm0.07$ & - \\
MATLAS-262 & $30.4$ & $7.04\pm0.10\,(7.32)$ & $47.9770$ & $-4.8933$ & $-6.82$ & $0.99$ & - & - & - & $5.67$ & - \\
MATLAS-347 & $12.3$ & $7.32\pm0.04\,(7.51)$ & $124.1349$ & $58.1908$ & $-4.81\pm0.04$ & $1.11\pm0.05$ & - & - & - & $5.07\pm0.08$ & - \\
MATLAS-365 & $30.8$ & $7.60\pm0.21\,(7.83)$ & $125.3902$ & $22.4681$ & $-9.16\pm0.06$ & $0.93\pm0.07$ & $6.34^*\pm0.78$ & $1.44\pm0.66$ & $0.43\pm0.14$ & $6.49\pm0.08$ & BOTH\\
MATLAS-368 & $30.8$ & $7.89\pm0.04\,(8.13)$ & $125.5730$ & $22.9083$ & $-11.15$ & $0.92$ & - & - & - & $7.28$ & - \\
MATLAS-405 & $28.0$ & $7.27\pm0.04\,(7.66)$ & $130.7773$ & $50.0788$ & $-7.69\pm0.20$ & $0.98\pm0.58$ & $6.20^*\pm1.06$ & $1.81\pm0.33$ & $0.33\pm0.07$ & $6.00\pm0.56$ & F606W\\
MATLAS-478 & $21.8$ & $7.73\pm0.01\,(8.07)$ & $138.2174$ & $59.9870$ & $-7.68\pm0.19$ & $0.81\pm0.49$ & $4.86^*\pm0.20$ & $0.58\pm0.25$ & $0.06\pm0.07$ & $5.69\pm0.47$ & F606W\\
MATLAS-524 & $27.0$ & $7.94\pm0.02\,(8.17)$ & $141.3540$ & $34.3998$ & $-7.94\pm0.18$ & $1.06\pm0.51$ & $5.07^*\pm0.44$ & $0.62\pm0.45$ & $0.28\pm0.13$ & $6.23\pm0.49$ & F606W\\
MATLAS-627 & $45.8$ & $7.13\pm0.05\,(7.55)$ & $151.0653$ & $59.0412$ & $-9.10$ & $0.97$ & - & - & - & $6.54$ & - \\
MATLAS-658 & $33.1$ & $7.05\pm0.34\,(7.21)$ & $154.4668$ & $22.3334$ & $-6.62\pm0.06$ & $1.57\pm0.07$ & - & - & - & $6.59\pm0.08$ & - \\
MATLAS-682 & $40.8$ & $6.48\pm0.10\,(6.90)$ & $155.5875$ & $12.9975$ & - & - & - & - & - & - & - \\
MATLAS-791 & $24.5$ & $7.11\pm0.07\,(7.43)$ & $162.8981$ & $28.2542$ & $-6.92\pm0.28$ & $0.80\pm0.73$ & $5.40^*\pm1.03$ & $1.21\pm0.62$ & $0.39\pm0.08$ & $5.37\pm0.70$ & F606W\\
MATLAS-976 & $26.3$ & $7.21\pm0.06\,(7.51)$ & $170.4676$ & $3.2406$ & $-7.90\pm0.22$ & $0.99\pm0.23$ & $8.24\pm0.71$ & $1.82\pm0.22$ & $0.15\pm0.12$ & $6.09\pm0.29$ & BOTH\\
MATLAS-984 & $33.1$ & $7.07\pm0.05\,(7.52)$ & $170.5743$ & $39.0395$ & $-7.55\pm0.27$ & $0.84\pm0.74$ & $6.06^*\pm0.81$ & $0.78\pm0.56$ & $0.28\pm0.14$ & $5.69\pm0.72$ & F606W\\
MATLAS-987 & $32.7$ & $7.39\pm0.05\,(7.61)$ & $170.6946$ & $38.4594$ & $-8.23\pm0.04$ & $0.80\pm0.50$ & - & - & - & $5.90\pm0.05$ & - \\
 &  &  & $170.6947$ & $38.4594$ & $-8.96\pm0.04$ & $0.89\pm0.05$ & - & - & - & $6.34\pm0.05$ & - \\
MATLAS-991 & $33.4$ & $6.45\pm0.12\,(6.84)$ & $170.7145$ & $56.9153$ & $-7.02\pm0.38$ & $0.68\pm0.83$ & $10.56\pm1.59$ & $0.55\pm0.25$ & $0.21\pm0.17$ & $5.21\pm0.82$ & F606W\\
MATLAS-1059 & $33.4$ & $7.74\pm0.02\,(8.14)$ & $171.9658$ & $56.855$ & $-10.67\pm0.05$ & $0.97\pm0.06$ & $10.92\pm1.11$ & $4.64\pm0.66$ & $0.20\pm0.09$ & $7.17\pm0.06$ & BOTH\\
MATLAS-1216 & $39.2$ & $7.19\pm0.07\,(7.56)$ & $183.7927$ & $7.3109$ & $-6.58\pm0.06$ & $1.38\pm0.07$ & - & - & - & $6.25\pm0.08$ & - \\
MATLAS-1262 & $31.5$ & $7.60\pm0.06\,(7.95)$ & $184.4666$ & $5.9889$ & $-9.31\pm0.05$ & $1.04\pm0.07$ & $8.01\pm0.42$ & $1.49\pm0.21$ & $0.27\pm0.08$ & $6.75\pm0.08$ & BOTH\\
MATLAS-1321 & $37.2$ & $7.80\pm0.02\,(8.17)$ & $185.0148$ & $5.7206$ & $-10.41\pm0.04$ & $0.91\pm0.05$ & $8.99^*\pm0.33$ & $5.31\pm0.03$ & $0.04\pm0.09$ & $6.96\pm0.06$ & BOTH\\
MATLAS-1332 & $35.8^a$ & $8.24\pm0.02\,(8.59)$ & $185.0674$ & $5.0392$ & $-11.88\pm0.03$ & $0.94\pm0.03$ & $11.42\pm0.21$ & $3.47\pm0.22$ & $0.33\pm0.06$ & $7.59\pm0.04$ & BOTH\\
MATLAS-1408 & $14.0^b$ & $7.93\pm0.02\,(8.25)$ & $190.2966$ & $-5.0979$ & $-10.05\pm0.07$ & $1.03\pm0.08$ & $4.86\pm0.19$ & $2.22\pm0.11$ & $0.09\pm0.08$ & $7.02\pm0.09$ & BOTH\\
MATLAS-1412 & $16.5$ & $6.87\pm0.04\,(7.33)$ & $190.3294$ & $2.1001$ & $-9.14\pm0.04$ & $0.83\pm0.04$ & $5.63\pm0.23$ & $2.35\pm0.31$ & $0.11\pm0.06$ & $6.31\pm0.05$ & BOTH\\
MATLAS-1437 & $16.5$ & $7.36\pm0.03\,(7.70)$ & $190.7285$ & $2.2520$ & $-9.14\pm0.08$ & $0.96\pm0.09$ & $7.67\pm0.25$ & $3.29\pm0.15$ & $0.33\pm0.10$ & $6.54\pm0.10$ & BOTH\\
MATLAS-1470 & $16.5$ & $7.72\pm0.03\,(8.08)$ & $191.2317$ & $2.2961$ & $-10.55\pm0.04$ & $0.90\pm0.04$ & $9.88\pm0.24$ & $3.45\pm0.13$ & $0.26\pm0.05$ & $7.00\pm0.05$ & BOTH\\
MATLAS-1485 & $13.9^c$ & $7.75\pm0.02\,(8.04)$ & $191.7060$ & $2.7134$ & $-10.68\pm0.02$ & $0.95\pm0.02$ & $6.21\pm0.16$ & $4.11\pm1.07$ & $0.07\pm0.03$ & $7.14\pm0.03$ & BOTH\\
MATLAS-1539 & $39.6$ & $7.83\pm0.03\,(8.15)$ & $202.4220$ & $46.6439$ & $-11.56\pm0.04$ & $1.03\pm0.04$ & $22.15\pm0.38$ & $2.47\pm0.46$ & $0.16\pm0.05$ & $7.62\pm0.05$ & BOTH\\
MATLAS-1545 & $39.6$ & $6.48\pm0.10\,(6.98)$ & $202.7980$ & $46.8160$ & $-7.96\pm0.90$ & $1.00\pm1.34$ & $7.80^*\pm1.34$ & $0.67\pm0.47$ & $0.21\pm0.10$ & $6.14\pm1.39$ & F606W\\
MATLAS-1558 & $31.5$ & $7.47\pm0.04\,(7.77)$ & $206.4765$ & $61.0970$ & $-9.66\pm0.72$ & $0.87\pm0.91$ & $5.14^*\pm0.00$ & $1.63\pm0.04$ & $0.19\pm0.16$ & $6.58\pm0.99$ & F814W\\
MATLAS-1577 & $31.5$ & $7.70\pm0.12\,(7.86)$ & $207.0560$ & $60.8855$ & $-10.22\pm0.1$ & $0.89\pm0.25$ & $6.90^*\pm0.96$ & $2.76\pm0.26$ & $0.15\pm0.12$ & $6.85\pm0.24$ & F606W\\
MATLAS-1616 & $30.0$ & $7.59\pm0.04\,(7.95)$ & $208.1219$ & $59.8761$ & $-9.41\pm0.07$ & $0.91\pm0.08$ & $8.12\pm0.22$ & $3.49\pm0.25$ & $0.24\pm0.11$ & $6.55\pm0.09$ & BOTH\\
MATLAS-1630 & $30.3$ & $7.79\pm0.05\,(8.04)$ & $208.2605$ & $60.1723$ & - & - & - & - & - & - & - \\
MATLAS-1662 & $37.1$ & $7.23\pm0.06\,(7.45)$ & $208.7525$ & $40.0551$ & $-7.91\pm0.27$ & $0.93\pm0.63$ & $10.31\pm2.43$ & $1.15\pm0.79$ & $0.55\pm0.12$ & $6.00\pm0.62$ & F606W\\
MATLAS-1667 & $30.0$ & $7.49\pm0.02\,(7.88)$ & $208.8191$ & $59.7469$ & $-7.67\pm0.24$ & $0.95\pm0.64$ & $6.18^*\pm0.95$ & $0.87\pm0.58$ & $0.38\pm0.14$ & $5.93\pm0.61$ & F606W\\
MATLAS-1740 & $25.8$ & $6.60\pm0.04\,(6.97)$ & $211.9086$ & $50.5049$ & $-6.71\pm0.95$ & $0.76\pm0.96$ & $5.35^*\pm0.75$ & $1.30\pm0.24$ & $0.54\pm0.22$ & $5.22\pm1.24$ & F814W\\
MATLAS-1779 & $38.8$ & $7.43\pm0.23\,(7.80)$ & $212.7724$ & $-5.2220$ & $-10.29\pm1.00$ & $0.86\pm1.34$ & $8.18^*\pm1.41$ & $3.80\pm0.19$ & $0.27\pm0.10$ & $6.83\pm1.43$ & F814W\\
MATLAS-1938 & $17.8^c$ & $8.41\pm0.01\,(8.75)$ & $225.1376$ & $2.2303$ & $-10.12\pm0.01$ & $0.91\pm0.02$ & - & - & - & $6.85\pm0.02
$ & - \\
MATLAS-1975 & $26.4$ & $7.02\pm0.34\,(7.40)$ & $225.8390$ & $1.4308$ & $-8.05\pm0.18$ & $0.89\pm0.53$ & $4.68^*\pm0.17$ & $1.12\pm0.21$ & $0.12\pm0.07$ & $5.98\pm0.51$ & F606W\\
MATLAS-1985 & $26.4$ & $7.44\pm0.23\,(7.64)$ & $226.0103$ & $1.3667$ & $-8.28\pm0.10$ & $0.79\pm0.12$ & $6.60\pm0.81$ & $1.69\pm0.26$ & $0.30\pm0.09$ & $5.90\pm0.13$ & BOTH\\
\bottomrule
\end{tabular}}
\begin{tablenotes}
\item $^*$ NSC R$_{\rm e}$ smaller than one pixel.
\item Distance measurement from $^a$\citet{Ann2015}, $^b$\citet{heesters2023}, and $^c$SDSS DR13 database \citep{Albareti2017,Blanton2017}. 
\end{tablenotes}
\end{table*}

\newpage
\clearpage
\section{NSC color profiles}
\label{AppendixB}

For each NSC successfully modeled with \textsc{galfit} and having R$_{\rm e,\rm NSC}>1$\,pix: we show the color profile within the 3R$_{\rm e,\rm NSC}$ region (left) based on the light profiles returned by \textsc{autoprof}, as well as the NSC $3\arcsec\times3\arcsec$ cutout (right) in the F606W (top) and F814W (bottom) filters with a 3R$_{\rm e,\rm NSC}$ radius circle. We ordered the NSCs by increasing M$_{\ast}^{\rm NSC}$.

\begin{figure*}
\centering
\begin{subfigure}{\textwidth}
\includegraphics[width=\linewidth]{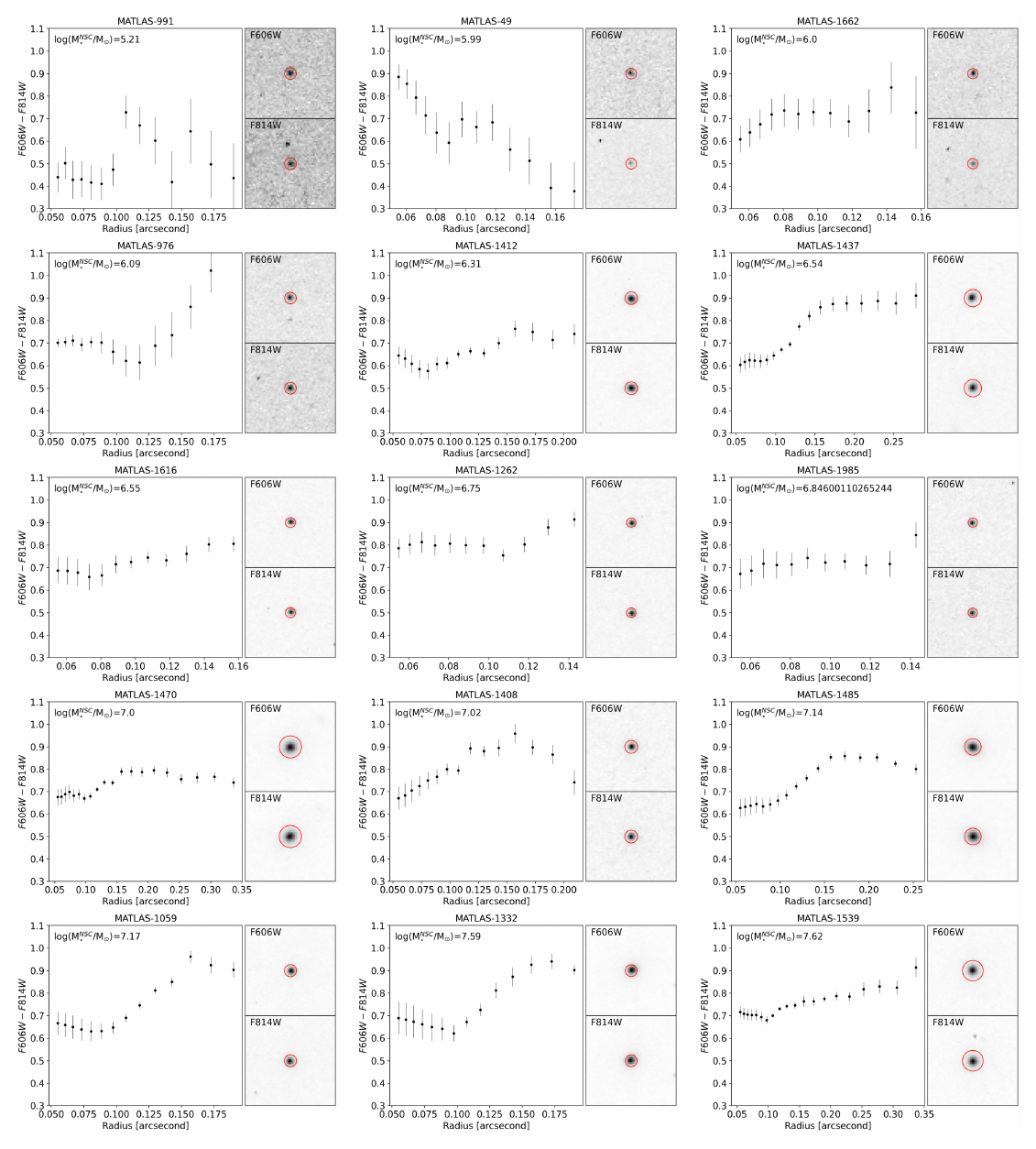}
\end{subfigure}
\caption{Color profiles of the 15 modeled NSCs with R$_{\rm e,\rm NSC}>1$\,pix.}
\label{fig:color-profiles}
\end{figure*}

\newpage
\clearpage
\section{GC properties}
\label{AppendixC}

Table \ref{tab:GCprop} presents the structural and photometric properties of the 142 marginally resolved GCs identified in 17 dwarf galaxies within a 25 Mpc distance \citep{Marleau2024}. For each GC, the table includes its RA/Dec coordinates, radial distance ($\delta_{GC}$) in terms of galaxy R$_{\rm e,\rm gal}$, absolute magnitude (M$_{V,GC}$), $(V-I)_0$ color, effective radius (R$_{\rm e,\rm GC}$), stellar mass (M$_{*}^{\rm GC}$), as well as the host galaxy with its distance. The last column indicates the band from which the parameters were derived. If the parameters are based on one band, 2R$_{\rm e,\rm GC}$ aperture photometry was applied in both bands to derive the color. When both bands were used, R$_{\rm e,\rm GC}$ corresponds to the average value of both bands. The GCs are sorted by increasing RA.

\begin{table*}
\small
\centering
\caption{\label{tab:GCprop}Structural and photometric properties of the 142 GCs.}
\begin{tabular}{cccccccccc}
\toprule
Host galaxy & Dist & RA & Dec & $\delta_{GC}$ & M$_V$ & $(V-I)_0$ & R$_{\rm e}$ & $log(\frac{M_{*}}{M_{\odot}})$ & Band\\
  & [Mpc] & [deg] & [deg] & [R$_{\rm e,\rm gal}$] & [mag] & [mag] & [pc] &  & \\
\toprule
MATLAS-177 & $22.4$ & $37.2001$ & $-1.3049$ & $1.44$ & $-8.13\pm0.62$ & $0.91\pm0.93$ & $4.37^*\pm0.43$ & $6.05\pm0.96$ & F606W\\
 &  & $37.2057$ & $-1.2995$ & $1.17$ & $-5.53\pm1.23$ & $0.98\pm1.84$ & $7.4\pm2.85$ & $5.12\pm1.91$ & F606W\\
MATLAS-347 & $12.3$ & $124.1262$ & $58.1888$ & $0.78$ & $-5.64\pm0.89$ & $0.98\pm0.98$ & $3.46\pm0.65$ & $5.17\pm1.14$ & F814W\\
 &  & $124.1355$ & $58.1899$ & $0.05$ & $-5.0\pm1.38$ & $0.93\pm2.0$ & $3.29\pm1.36$ & $4.82\pm2.09$ & F606W\\
MATLAS-478 & $21.8$ & $138.222$ & $59.9828$ & $0.84$ & $-5.25\pm1.02$ & $1.08\pm1.66$ & $5.73\pm3.19$ & $5.19\pm1.69$ & F606W\\
 &  & $138.2252$ & $59.9852$ & $0.78$ & $-5.75\pm1.84$ & $0.01\pm3.4$ & $1.21^*\pm0.01$ & $3.54\pm3.4$ & F606W\\
 &  & $138.2283$ & $59.9848$ & $1.06$ & $-5.3\pm1.74$ & $0.49\pm2.59$ & $4.72^*\pm3.69$ & $4.19\pm2.69$ & F606W\\
 &  & $138.2317$ & $59.9952$ & $1.91$ & $-5.37\pm1.18$ & $1.6\pm1.41$ & $6.87\pm1.62$ & $6.13\pm1.57$ & BOTH\\
MATLAS-799 & $24.5$ & $162.9928$ & $28.3567$ & $1.16$ & $-5.26\pm1.72$ & $0.76\pm2.48$ & $8.68\pm4.95$ & $4.64\pm2.59$ & F606W\\
 &  & $162.9946$ & $28.3588$ & $0.68$ & $-5.91\pm1.4$ & $1.19\pm1.66$ & $10.96\pm3.07$ & $5.64\pm1.86$ & F814W\\
 &  & $162.9946$ & $28.3549$ & $1.23$ & $-7.14\pm1.36$ & $1.03\pm1.45$ & $5.67^*\pm1.16$ & $5.86\pm1.74$ & BOTH\\
 &  & $162.9948$ & $28.3622$ & $0.72$ & $-5.45\pm1.13$ & $1.0\pm1.7$ & $13.95\pm3.79$ & $5.13\pm1.77$ & F606W\\
 &  & $162.9958$ & $28.3547$ & $1.16$ & $-6.19\pm0.86$ & $1.35\pm0.86$ & $10.7\pm0.0$ & $6.04\pm1.14$ & BOTH\\
 &  & $162.9963$ & $28.359$ & $0.39$ & $-5.73\pm1.12$ & $0.58\pm1.7$ & $13.07\pm4.92$ & $4.51\pm1.76$ & F606W\\
 &  & $162.9976$ & $28.3645$ & $0.89$ & $-8.24\pm0.43$ & $0.9\pm0.69$ & $5.42^*\pm0.1$ & $6.07\pm0.71$ & F606W\\
 &  & $162.9977$ & $28.3638$ & $0.76$ & $-6.18\pm1.49$ & $0.94\pm2.12$ & $8.32\pm2.47$ & $5.32\pm2.22$ & F606W\\
 &  & $162.9983$ & $28.3665$ & $1.29$ & $-6.87\pm1.03$ & $1.2\pm1.27$ & $10.35\pm1.81$ & $6.04\pm1.4$ & F814W\\
 &  & $162.999$ & $28.3622$ & $0.46$ & $-7.62\pm0.4$ & $0.92\pm0.5$ & $4.62^*\pm0.07$ & $5.85\pm0.55$ & BOTH\\
 &  & $162.9994$ & $28.3651$ & $1.05$ & $-5.3\pm1.5$ & $1.2\pm1.74$ & $6.5\pm4.65$ & $5.41\pm1.97$ & F814W\\
 &  & $163.0018$ & $28.361$ & $0.7$ & $-8.25\pm0.46$ & $0.75\pm0.49$ & $5.78^*\pm0.03$ & $5.81\pm0.59$ & BOTH\\
 &  & $163.0036$ & $28.3636$ & $1.22$ & $-6.79\pm1.2$ & $0.62\pm1.55$ & $9.75\pm1.12$ & $5.01\pm1.68$ & BOTH\\
 &  & $163.0041$ & $28.3662$ & $1.65$ & $-8.16\pm0.44$ & $0.5\pm0.65$ & $6.54\pm0.01$ & $5.35\pm0.68$ & BOTH\\
 &  & $163.0056$ & $28.3543$ & $1.8$ & $-5.83\pm1.58$ & $0.83\pm1.9$ & $9.34\pm3.29$ & $4.99\pm2.11$ & F814W\\
MATLAS-898 & $19.8$ & $169.2516$ & $18.3045$ & $1.17$ & $-7.36\pm0.39$ & $1.27\pm0.69$ & $4.05^*\pm0.08$ & $6.36\pm0.69$ & F606W\\
 &  & $169.2532$ & $18.3034$ & $0.79$ & $-5.28\pm1.1$ & $0.56\pm1.68$ & $6.41\pm3.54$ & $4.3\pm1.73$ & F606W\\
 &  & $169.2589$ & $18.3016$ & $0.45$ & $-7.12\pm0.82$ & $1.13\pm1.19$ & $5.27\pm0.95$ & $6.02\pm1.24$ & F606W\\
 &  & $169.2596$ & $18.3083$ & $1.3$ & $-5.92\pm1.2$ & $2.14\pm1.22$ & $2.53^*\pm0.11$ & $7.29\pm1.55$ & F814W\\
MATLAS-1154 & $24.6$ & $180.3977$ & $61.8333$ & $0.97$ & $-5.63\pm1.79$ & $1.6\pm1.95$ & $11.22\pm1.88$ & $6.25\pm2.29$ & BOTH\\
 &  & $180.3982$ & $61.8369$ & $0.32$ & $-4.47\pm2.5$ & $1.12\pm2.9$ & $3.78^*\pm4.44$ & $4.95\pm3.28$ & F814W\\
 &  & $180.4035$ & $61.8373$ & $0.61$ & $-7.23\pm1.18$ & $0.95\pm1.32$ & $9.71\pm1.22$ & $5.76\pm1.53$ & BOTH\\
MATLAS-1225 & $19.1$ & $184.0979$ & $27.7433$ & $1.18$ & $-6.37\pm1.15$ & $1.5\pm1.39$ & $5.87\pm1.66$ & $6.36\pm1.54$ & BOTH\\
 &  & $184.1043$ & $27.7405$ & $0.63$ & $-6.19\pm1.18$ & $0.84\pm1.39$ & $6.76\pm1.12$ & $5.14\pm1.56$ & F814W\\
 &  & $184.1046$ & $27.7434$ & $0.06$ & $-5.27\pm1.23$ & $0.66\pm2.01$ & $3.94^*\pm0.01$ & $4.47\pm2.05$ & F606W\\
MATLAS-1400 & $17.0^a$ & $190.2156$ & $7.939$ & $1.94$ & $-4.89\pm1.5$ & $1.38\pm2.19$ & $5.38\pm2.86$ & $5.57\pm2.28$ & F606W\\
 &  & $190.2168$ & $7.9304$ & $1.0$ & $-5.45\pm1.05$ & $0.67\pm1.6$ & $4.78\pm0.55$ & $4.55\pm1.65$ & F606W\\
 &  & $190.2181$ & $7.9241$ & $1.98$ & $-5.75\pm1.4$ & $0.81\pm1.69$ & $4.72\pm1.52$ & $4.91\pm1.88$ & F814W\\
MATLAS-1412 & $16.5$ & $190.3264$ & $2.096$ & $1.19$ & $-7.2\pm0.07$ & $1.49\pm0.29$ & $5.69\pm0.0$ & $6.68\pm0.28$ & BOTH\\
 &  & $190.3265$ & $2.0963$ & $1.12$ & $-6.35\pm0.83$ & $1.47\pm1.03$ & $11.76\pm1.46$ & $6.31\pm1.13$ & F814W\\
 &  & $190.3293$ & $2.094$ & $1.43$ & $-7.22\pm0.36$ & $1.01\pm0.43$ & $4.87\pm0.53$ & $5.86\pm0.48$ & BOTH\\
 &  & $190.3312$ & $2.1078$ & $1.82$ & $-6.99\pm0.83$ & $0.91\pm0.86$ & $2.75^*\pm0.01$ & $5.59\pm1.05$ & F814W\\
 &  & $190.3335$ & $2.0948$ & $1.58$ & $-5.68\pm1.27$ & $0.89\pm1.82$ & $4.96\pm2.15$ & $5.04\pm1.91$ & F606W\\
 &  & $190.3353$ & $2.1$ & $1.38$ & $-7.68\pm0.26$ & $0.94\pm0.33$ & $4.31\pm0.05$ & $5.91\pm0.36$ & BOTH\\
MATLAS-1437 & $16.5$ & $190.7283$ & $2.2545$ & $0.39$ & $-6.22\pm0.83$ & $1.24\pm0.87$ & $3.47^*\pm0.11$ & $5.85\pm1.06$ & F814W\\
MATLAS-1470 & $16.5$ & $191.2299$ & $2.2969$ & $0.43$ & $-6.69\pm1.03$ & $0.6\pm1.16$ & $6.1\pm1.07$ & $4.94\pm1.33$ & BOTH\\
 &  & $191.2311$ & $2.2954$ & $0.18$ & $-8.29\pm0.18$ & $1.0\pm0.22$ & $4.59\pm0.18$ & $6.26\pm0.24$ & BOTH\\
 &  & $191.2312$ & $2.2957$ & $0.11$ & $-5.62\pm1.29$ & $0.95\pm1.58$ & $6.15\pm1.75$ & $5.12\pm1.75$ & F814W\\
 &  & $191.2327$ & $2.2994$ & $0.87$ & $-7.39\pm0.32$ & $0.74\pm0.5$ & $3.28^*\pm0.03$ & $5.46\pm0.51$ & BOTH\\
 &  & $191.2332$ & $2.2996$ & $0.96$ & $-8.22\pm0.23$ & $0.95\pm0.27$ & $5.58\pm0.21$ & $6.15\pm0.31$ & BOTH\\
 &  & $191.2357$ & $2.2939$ & $1.16$ & $-7.8\pm0.31$ & $0.86\pm0.53$ & $3.95^*\pm0.07$ & $5.82\pm0.54$ & F606W\\
\bottomrule
\end{tabular}
\begin{tablenotes}
\item $^*$ GC R$_{\rm e}$ smaller than one pixel.
\item Distance measurement from $^a$\citet{heesters2023}, $^b$SDSS DR13 database \citep{Albareti2017,Blanton2017}, and $^c$\citet{Mueller2021}. 
\end{tablenotes}
\end{table*}

\begin{table*}
\ContinuedFloat
\small
\caption{continued.}
\begin{tabular}{cccccccccc}
\toprule
Host galaxy & Dist & RA & Dec & $\delta_{GC}$ & M$_V$ & $(V-I)_0$ & R$_{\rm e}$ & $log(\frac{M_{*}}{M_{\odot}})$ & Band\\
  & [Mpc] & [deg] & [deg] & [R$_{\rm e,\rm gal}$] & [mag] & [mag] & [pc] &  & \\
\toprule
MATLAS-1485 & $13.9^b$ & $191.7012$ & $2.7066$ & $1.89$ & $-7.4\pm0.57$ & $0.8\pm0.65$ & $7.7\pm1.72$ & $5.57\pm0.74$ & F814W\\
 &  & $191.7015$ & $2.7096$ & $1.33$ & $-5.95\pm1.02$ & $1.46\pm1.5$ & $3.45\pm0.94$ & $6.13\pm1.56$ & F606W\\
 &  & $191.7023$ & $2.7137$ & $0.83$ & $-6.72\pm1.24$ & $0.86\pm1.4$ & $5.06\pm0.67$ & $5.39\pm1.61$ & BOTH\\
 &  & $191.7036$ & $2.7191$ & $1.37$ & $-7.6\pm0.34$ & $1.03\pm0.4$ & $3.43\pm0.01$ & $6.04\pm0.45$ & BOTH\\
 &  & $191.7076$ & $2.7144$ & $0.42$ & $-6.4\pm0.97$ & $0.74\pm1.01$ & $2.73^*\pm0.67$ & $5.06\pm1.23$ & F814W\\
 &  & $191.7085$ & $2.7056$ & $1.86$ & $-6.16\pm1.21$ & $0.75\pm1.51$ & $6.3\pm1.18$ & $4.97\pm1.66$ & F814W\\
 &  & $191.709$ & $2.7185$ & $1.32$ & $-9.53\pm0.16$ & $0.91\pm0.18$ & $6.88\pm0.12$ & $6.61\pm0.21$ & BOTH\\
 &  & $191.711$ & $2.7057$ & $2.08$ & $-7.86\pm0.28$ & $0.99\pm0.31$ & $2.38^*\pm0.02$ & $6.08\pm0.36$ & BOTH\\
 &  & $191.7126$ & $2.7182$ & $1.83$ & $-6.44\pm1.15$ & $0.89\pm1.61$ & $5.5\pm1.14$ & $5.33\pm1.7$ & F606W\\
 &  & $191.7128$ & $2.7027$ & $2.86$ & $-8.39\pm0.26$ & $0.92\pm0.29$ & $7.36\pm0.22$ & $6.16\pm0.33$ & BOTH\\
 &  & $191.7163$ & $2.7196$ & $2.7$ & $-6.71\pm0.79$ & $0.89\pm1.13$ & $7.08\pm0.43$ & $5.44\pm1.19$ & F606W\\
MATLAS-1907 & $24.2$ & $217.7099$ & $3.2053$ & $0.69$ & $-5.63\pm1.37$ & $0.32\pm2.12$ & $4.97^*\pm4.25$ & $4.02\pm2.18$ & F606W\\
MATLAS-1907 & $24.2$ & $217.7099$ & $3.2053$ & $0.69$ & $-5.63\pm1.37$ & $0.32\pm2.12$ & $4.97^*\pm4.25$ & $4.02\pm2.18$ & F606W\\
 &  & $217.7148$ & $3.2122$ & $1.71$ & $-5.78\pm1.04$ & $0.96\pm1.6$ & $7.87\pm3.12$ & $5.19\pm1.65$ & F606W\\
 &  & $217.715$ & $3.2043$ & $0.99$ & $-6.33\pm1.61$ & $0.93\pm1.86$ & $9.49\pm2.39$ & $5.37\pm2.11$ & BOTH\\
MATLAS-1938 & $17.8^b$ & $225.132$ & $2.2307$ & $1.58$ & $-9.54\pm0.31$ & $0.89\pm0.52$ & $3.71^*\pm0.05$ & $6.56\pm0.52$ & F606W\\
 &  & $225.1333$ & $2.2249$ & $1.96$ & $-8.58\pm0.57$ & $0.86\pm0.88$ & $4.05^*\pm0.04$ & $6.14\pm0.91$ & F606W\\
 &  & $225.1343$ & $2.229$ & $1.01$ & $-9.14\pm0.32$ & $0.85\pm0.36$ & $4.25^*\pm0.06$ & $6.35\pm0.42$ & BOTH\\
 &  & $225.1345$ & $2.2356$ & $1.69$ & $-9.1\pm0.41$ & $0.9\pm0.67$ & $3.77^*\pm0.04$ & $6.41\pm0.68$ & F606W\\
 &  & $225.1348$ & $2.2336$ & $1.18$ & $-8.44\pm0.5$ & $0.94\pm0.81$ & $5.04\pm0.01$ & $6.21\pm0.83$ & F606W\\
 &  & $225.1353$ & $2.2332$ & $1.0$ & $-7.43\pm0.92$ & $0.84\pm1.42$ & $3.46^*\pm0.87$ & $5.64\pm1.46$ & F606W\\
 &  & $225.1358$ & $2.2363$ & $1.72$ & $-9.64\pm0.55$ & $0.94\pm0.78$ & $4.89\pm0.44$ & $6.71\pm0.82$ & F606W\\
 &  & $225.136$ & $2.2313$ & $0.51$ & $-8.83\pm0.64$ & $0.98\pm0.66$ & $3.65^*\pm0.01$ & $6.45\pm0.82$ & F814W\\
 &  & $225.1363$ & $2.2254$ & $1.44$ & $-8.53\pm0.76$ & $0.37\pm0.96$ & $5.4\pm0.7$ & $5.26\pm1.05$ & BOTH\\
 &  & $225.1371$ & $2.2264$ & $1.14$ & $-8.15\pm0.81$ & $0.71\pm1.27$ & $3.05^*\pm0.01$ & $5.71\pm1.3$ & F606W\\
 &  & $225.1371$ & $2.2308$ & $0.17$ & $-8.0\pm0.17$ & $0.87\pm0.68$ & $3.25^*\pm0.0$ & $5.92\pm0.64$ & F606W\\
 &  & $225.1373$ & $2.225$ & $1.53$ & $-7.56\pm1.03$ & $0.86\pm1.55$ & $3.13^*\pm1.89$ & $5.74\pm1.6$ & F606W\\
 &  & $225.1376$ & $2.2332$ & $0.76$ & $-8.76\pm0.72$ & $0.83\pm1.07$ & $2.93^*\pm1.03$ & $6.15\pm1.12$ & F606W\\
 &  & $225.1378$ & $2.2308$ & $0.12$ & $-9.02\pm0.45$ & $0.92\pm0.7$ & $4.49\pm0.04$ & $6.41\pm0.72$ & F606W\\
 &  & $225.1379$ & $2.2309$ & $0.15$ & $-8.07\pm0.81$ & $0.75\pm1.22$ & $3.3^*\pm1.52$ & $5.75\pm1.27$ & F606W\\
 &  & $225.1382$ & $2.2294$ & $0.34$ & $-8.07\pm1.17$ & $0.78\pm1.67$ & $3.44^*\pm1.55$ & $5.8\pm1.75$ & F606W\\
 &  & $225.1384$ & $2.2307$ & $0.23$ & $-8.17\pm0.86$ & $1.03\pm0.9$ & $3.19^*\pm0.01$ & $6.27\pm1.09$ & F814W\\
 &  & $225.1384$ & $2.2327$ & $0.67$ & $-7.25\pm0.99$ & $0.85\pm1.46$ & $4.77\pm1.96$ & $5.58\pm1.52$ & F606W\\
 &  & $225.1388$ & $2.2313$ & $0.4$ & $-9.1\pm0.28$ & $0.7\pm0.34$ & $3.84^*\pm0.01$ & $6.07\pm0.38$ & BOTH\\
 &  & $225.1393$ & $2.2283$ & $0.76$ & $-7.44\pm1.17$ & $0.93\pm1.68$ & $4.86\pm1.89$ & $5.81\pm1.76$ & F606W\\
 &  & $225.1394$ & $2.2279$ & $0.86$ & $-9.31\pm0.43$ & $0.86\pm0.47$ & $4.23^*\pm0.0$ & $6.43\pm0.55$ & BOTH\\
 &  & $225.1397$ & $2.231$ & $0.61$ & $-6.41\pm1.79$ & $1.05\pm2.53$ & $7.82\pm2.76$ & $5.6\pm2.67$ & F606W\\
 &  & $225.14$ & $2.2302$ & $0.66$ & $-8.92\pm0.69$ & $0.93\pm0.99$ & $4.36\pm0.8$ & $6.39\pm1.04$ & F606W\\
 &  & $225.1407$ & $2.2284$ & $1.04$ & $-6.13\pm1.3$ & $1.05\pm1.92$ & $5.67\pm3.43$ & $5.49\pm2.0$ & F606W\\
 &  & $225.1408$ & $2.2343$ & $1.39$ & $-6.91\pm1.24$ & $1.32\pm1.33$ & $3.96^*\pm0.41$ & $6.27\pm1.58$ & BOTH\\
 &  & $225.1423$ & $2.2318$ & $1.35$ & $-8.69\pm0.66$ & $0.86\pm0.68$ & $3.86^*\pm0.15$ & $6.18\pm0.85$ & F814W\\
MATLAS-2019 & $20.2^c$ & $226.3262$ & $1.8172$ & $1.88$ & $-7.72\pm1.37$ & $0.4\pm1.48$ & $4.51^*\pm1.06$ & $4.99\pm1.75$ & BOTH\\
 &  & $226.3266$ & $1.8172$ & $1.82$ & $-5.22\pm1.51$ & $1.27\pm2.26$ & $5.19\pm3.85$ & $5.51\pm2.35$ & F606W\\
 &  & $226.3268$ & $1.8098$ & $1.63$ & $-6.86\pm1.43$ & $0.73\pm1.66$ & $5.71\pm1.38$ & $5.22\pm1.87$ & BOTH\\
 &  & $226.3288$ & $1.8111$ & $1.14$ & $-7.09\pm0.89$ & $1.08\pm1.21$ & $6.24\pm1.13$ & $5.92\pm1.29$ & BOTH\\
 &  & $226.3298$ & $1.812$ & $0.88$ & $-6.53\pm1.05$ & $1.05\pm1.54$ & $6.58\pm2.21$ & $5.64\pm1.61$ & F606W\\
 &  & $226.33$ & $1.8115$ & $0.87$ & $-8.17\pm0.61$ & $0.97\pm0.89$ & $5.39\pm0.0$ & $6.16\pm0.93$ & F606W\\
 &  & $226.3301$ & $1.8142$ & $0.88$ & $-7.58\pm0.89$ & $1.03\pm1.32$ & $3.88^*\pm0.84$ & $6.03\pm1.37$ & F606W\\
 &  & $226.3301$ & $1.8101$ & $0.97$ & $-8.07\pm0.7$ & $0.88\pm0.8$ & $7.22\pm0.0$ & $5.97\pm0.91$ & F814W\\
 &  & $226.3307$ & $1.8124$ & $0.69$ & $-7.1\pm1.18$ & $0.78\pm1.78$ & $2.73^*\pm1.85$ & $5.42\pm1.84$ & F606W\\
 &  & $226.3311$ & $1.8106$ & $0.73$ & $-7.7\pm0.67$ & $0.86\pm0.77$ & $5.98\pm0.78$ & $5.78\pm0.88$ & BOTH\\
 &  & $226.3312$ & $1.8134$ & $0.61$ & $-7.25\pm1.19$ & $1.16\pm1.71$ & $4.56^*\pm1.43$ & $6.13\pm1.79$ & F606W\\
 &  & $226.3314$ & $1.8125$ & $0.54$ & $-6.61\pm1.09$ & $0.99\pm1.43$ & $6.09\pm1.18$ & $5.58\pm1.54$ & BOTH\\
 &  & $226.3314$ & $1.8151$ & $0.75$ & $-8.62\pm0.41$ & $0.77\pm0.45$ & $5.15\pm0.03$ & $5.99\pm0.52$ & BOTH\\
 &  & $226.3319$ & $1.8111$ & $0.54$ & $-7.5\pm0.95$ & $1.05\pm1.39$ & $4.23^*\pm0.0$ & $6.03\pm1.45$ & F606W\\
 &  & $226.3322$ & $1.8108$ & $0.54$ & $-6.08\pm1.4$ & $0.65\pm2.04$ & $5.52\pm3.24$ & $4.78\pm2.13$ & F606W\\
 &  & $226.3327$ & $1.8139$ & $0.37$ & $-6.51\pm1.41$ & $1.02\pm2.05$ & $4.07^*\pm2.33$ & $5.59\pm2.14$ & F606W\\
 &  & $226.3328$ & $1.8095$ & $0.71$ & $-6.59\pm1.24$ & $1.15\pm1.39$ & $3.97^*\pm0.51$ & $5.85\pm1.61$ & F814W\\
 &  & $226.3329$ & $1.8127$ & $0.22$ & $-7.24\pm0.92$ & $1.01\pm1.39$ & $3.9^*\pm1.5$ & $5.86\pm1.44$ & F606W\\
 &  & $226.3333$ & $1.8175$ & $1.03$ & $-6.56\pm1.23$ & $0.95\pm1.84$ & $3.95^*\pm1.59$ & $5.49\pm1.91$ & F606W\\
 &  & $226.3335$ & $1.8111$ & $0.35$ & $-8.76\pm0.41$ & $0.93\pm0.63$ & $5.11\pm0.04$ & $6.34\pm0.65$ & F606W\\
 &  & $226.3337$ & $1.8152$ & $0.54$ & $-6.41\pm1.52$ & $1.07\pm2.2$ & $4.16^*\pm2.75$ & $5.63\pm2.3$ & F606W\\
\bottomrule
\end{tabular}
\begin{tablenotes}
\item $^*$ GC R$_{\rm e}$ smaller than one pixel.
\item Distance measurement from $^a$\citet{heesters2023}, $^b$SDSS DR13 database \citep{Albareti2017,Blanton2017}, and $^c$\citet{Mueller2021}. 
\end{tablenotes}
\end{table*}

\begin{table*}
\ContinuedFloat
\small
\caption{continued.}
\begin{tabular}{cccccccccc}
\toprule
Host galaxy & Dist & RA & Dec & $\delta_{GC}$ & M$_V$ & $(V-I)_0$ & R$_{\rm e}$ & $log(\frac{M_{*}}{M_{\odot}})$ & Band\\
  & [Mpc] & [deg] & [deg] & [R$_{\rm e,\rm gal}$] & [mag] & [mag] & [pc] &  & \\
\toprule
 &  & $226.3337$ & $1.8157$ & $0.64$ & $-7.18\pm0.93$ & $0.83\pm1.38$ & $5.24\pm0.0$ & $5.52\pm1.43$ & F606W\\
 &  & $226.3339$ & $1.8106$ & $0.43$ & $-8.67\pm0.55$ & $1.0\pm0.56$ & $4.16^*\pm0.37$ & $6.42\pm0.7$ & F814W\\
 &  & $226.3339$ & $1.8124$ & $0.05$ & $-9.54\pm0.05$ & $0.91\pm0.06$ & $5.78\pm1.83$ & $6.6\pm0.07$ & BOTH\\
 &  & $226.3341$ & $1.811$ & $0.35$ & $-6.56\pm1.36$ & $0.94\pm1.51$ & $3.44^*\pm1.14$ & $5.48\pm1.75$ & BOTH\\
 &  & $226.3341$ & $1.8102$ & $0.53$ & $-8.05\pm0.77$ & $0.96\pm1.11$ & $5.23\pm0.03$ & $6.09\pm1.16$ & F606W\\
 &  & $226.3346$ & $1.813$ & $0.14$ & $-9.92\pm0.05$ & $0.93\pm0.07$ & $7.85\pm0.18$ & $6.79\pm0.07$ & BOTH\\ 
 &  & $226.3349$ & $1.81$ & $0.59$ & $-6.46\pm1.73$ & $0.7\pm2.45$ & $6.33\pm0.02$ & $5.02\pm2.58$ & F606W\\
 &  & $226.3351$ & $1.8109$ & $0.44$ & $-7.07\pm0.82$ & $1.49\pm0.98$ & $5.79\pm0.01$ & $6.63\pm1.09$ & BOTH\\
 &  & $226.3352$ & $1.8137$ & $0.34$ & $-8.59\pm0.4$ & $1.22\pm0.45$ & $4.61^*\pm0.0$ & $6.77\pm0.52$ & BOTH\\
 &  & $226.3354$ & $1.822$ & $1.98$ & $-7.25\pm1.05$ & $0.95\pm1.61$ & $2.76^*\pm0.01$ & $5.77\pm1.66$ & F606W\\
 &  & $226.3356$ & $1.8126$ & $0.34$ & $-8.01\pm0.92$ & $1.04\pm1.34$ & $3.57^*\pm0.02$ & $6.23\pm1.4$ & F606W\\
 &  & $226.3357$ & $1.8116$ & $0.42$ & $-8.89\pm0.29$ & $0.93\pm0.33$ & $5.01\pm0.01$ & $6.39\pm0.38$ & BOTH\\
 &  & $226.3358$ & $1.8136$ & $0.44$ & $-7.78\pm0.95$ & $1.42\pm0.98$ & $3.45^*\pm0.0$ & $6.78\pm1.21$ & BOTH\\
 &  & $226.3365$ & $1.8168$ & $1.02$ & $-5.49\pm1.66$ & $0.93\pm2.47$ & $4.69^*\pm2.88$ & $5.02\pm2.56$ & F606W\\
 &  & $226.3365$ & $1.8152$ & $0.76$ & $-8.55\pm0.32$ & $1.04\pm0.61$ & $3.62^*\pm0.05$ & $6.44\pm0.61$ & F606W\\
 &  & $226.3366$ & $1.8175$ & $1.15$ & $-8.85\pm0.37$ & $0.97\pm0.63$ & $3.36^*\pm0.02$ & $6.43\pm0.64$ & F606W\\
 &  & $226.337$ & $1.812$ & $0.65$ & $-6.01\pm1.52$ & $0.91\pm2.19$ & $7.47\pm1.95$ & $5.2\pm2.29$ & F606W\\
 &  & $226.3376$ & $1.8173$ & $1.24$ & $-7.28\pm0.95$ & $0.87\pm1.17$ & $4.29^*\pm1.7$ & $5.64\pm1.29$ & BOTH\\
 &  & $226.3384$ & $1.8183$ & $1.51$ & $-6.86\pm1.09$ & $1.34\pm1.19$ & $6.13\pm1.06$ & $6.28\pm1.39$ & BOTH\\
 &  & $226.3402$ & $1.8171$ & $1.61$ & $-6.79\pm1.46$ & $0.8\pm2.09$ & $4.03^*\pm3.15$ & $5.31\pm2.19$ & F606W\\
 &  & $226.3406$ & $1.8091$ & $1.59$ & $-7.36\pm1.21$ & $0.51\pm1.4$ & $5.55\pm1.12$ & $5.05\pm1.59$ & BOTH\\
MATLAS-2021 & $21.8$ & $226.3328$ & $2.1665$ & $1.86$ & $-5.99\pm1.09$ & $0.89\pm1.63$ & $6.65\pm2.03$ & $5.16\pm1.7$ & F606W\\
 &  & $226.3391$ & $2.1645$ & $1.57$ & $-6.43\pm0.84$ & $0.88\pm1.38$ & $3.73^*\pm2.37$ & $5.32\pm1.4$ & F606W\\
 &  & $226.3442$ & $2.1773$ & $1.33$ & $-6.47\pm1.01$ & $1.45\pm1.4$ & $7.42\pm1.61$ & $6.32\pm1.48$ & BOTH\\
 &  & $226.3464$ & $2.1784$ & $1.8$ & $-5.43\pm1.71$ & $0.47\pm2.48$ & $5.62\pm4.27$ & $4.2\pm2.59$ & F606W\\
MATLAS-2176 & $23.2$ & $345.4315$ & $16.4581$ & $1.8$ & $-7.0\pm0.9$ & $1.01\pm1.39$ & $4.07^*\pm1.71$ & $5.77\pm1.43$ & F606W\\
 &  & $345.4335$ & $16.4514$ & $1.8$ & $-5.98\pm1.08$ & $0.96\pm1.61$ & $7.63\pm3.67$ & $5.27\pm1.67$ & F606W\\
 &  & $345.4346$ & $16.4589$ & $1.18$ & $-5.68\pm0.96$ & $1.09\pm1.5$ & $7.06\pm3.92$ & $5.38\pm1.54$ & F606W\\
 &  & $345.4358$ & $16.4552$ & $0.59$ & $-5.35\pm1.29$ & $0.83\pm2.05$ & $5.94\pm2.9$ & $4.79\pm2.1$ & F606W\\
 &  & $345.4366$ & $16.4551$ & $0.43$ & $-6.08\pm2.06$ & $0.82\pm3.14$ & $2.02^*\pm1.63$ & $5.07\pm3.24$ & F606W\\
 &  & $345.4383$ & $16.4577$ & $0.47$ & $-7.42\pm0.78$ & $1.11\pm1.2$ & $4.24^*\pm0.01$ & $6.11\pm1.24$ & F606W\\
 &  & $345.4402$ & $16.4598$ & $1.28$ & $-5.62\pm1.16$ & $0.84\pm1.73$ & $11.49\pm4.22$ & $4.91\pm1.79$ & F606W\\
\bottomrule
\end{tabular}
\begin{tablenotes}
\item $^*$ GC R$_{\rm e}$ smaller than one pixel.
\end{tablenotes}
\end{table*}

\end{appendix}

\end{document}